# Hydroacoustic Attenuation and Amplification by Turbulence

Kai-Xin Hu[a,b,1], Yue-Jin Hu[c]

[a]Zhejiang Provincial Engineering Research Center for the Safety of Pressure Vessel and Pipeline, Ningbo University, Ningbo, Zhejiang 315211, China

[b]Key Laboratory of Impact and Safety Engineering (Ningbo University), Ministry of Education, Ningbo, Zhejiang 315211, China

[c]Ningbo Jiangbei District People's Government, Ningbo, Zhejiang, 315800, China

## ABSTRACT

Acoustic waves propagating through fluid media are significantly influenced by turbulence. This paper experimentally investigates the influence of water turbulence on the propagation characteristics of acoustic waves. It shows that acoustic waves can be obviously attenuated or amplified at frequencies far exceeding the turbulent fluctuation frequency, with no spectral broadening. The amplification factor depends on the wave frequency rather than its amplitude. The study covers two flow conditions: pipe flow and free jet, driven by either a pump or hydraulic head difference, while the wave propagation directions both parallel and perpendicular to the mean flow are considered.Turbulent fluctuations without mean motion can still alter the wave amplitude, while laminar flow has no effect on acoustic signals. Comparison with conventional theories and experiments indicates that mechanisms such as bubbles, resonance, scattering, multipath interference or viscous dissipation cannot explain the observed phenomena. This indicates that there exists an incompletely understood new mechanism in the interaction between turbulence and acoustic waves.

[1]Corresponding author, Email: hukaixin@nbu.edu.cn

# I. INTRODUCTION

When acoustic waves propagate through a fluid medium, their amplitude, frequency, and direction of propagation can be altered by turbulence. This phenomenon is of great significance for understanding the laws of acoustic wave propagation and the characteristics of turbulent motion, making the interaction between acoustic waves and turbulence a subject of enduring research interest.

The earliest studies on this topic trace back to Lighthill (1952) in his pioneering work on aerodynamic sound. This issue represents not only a fundamental theoretical challenge in fluid dynamics but also holds substantial applied value in fields such as aircraft noise and hydroacoustics. Consequently, subsequent research has yielded significant theoretical and experimental progress in related areas.

First, acoustic waves can be scattered by turbulence. When a sound wave passes through a turbulent region with concentrated vorticity, spatial inhomogeneities in sound speed cause wave scattering. This problem was first independently studied by Lighthill (1953) and Kraichnan (1953), and later extended by multiple researchers (Baerg and Schwarz, 1966; Brown and Clifford, 1976). A key characteristic of sound wave scattering by turbulence is the spectral broadening. Campos (1978a, b) conducted a theoretical study on the spectral broadening of sound wave through turbulent shear layers and compared it with the experimental results relevant to the study of aircraft noise. Korman and Beyer (1980,1988) experimentally examined the scattering of sound by turbulence in water, confirming the characteristics of spectral broadening. In this experiment, a single monochromatic sound wave is scattered in a confined region of turbulence which is created by a submerged water jet.

Second, turbulence can absorb acoustic waves. In studies of sound scattering by turbulence, researchers typically assume that turbulence remains unaffected by the acoustic wave, with sound energy merely being redistributed in all directions. This assumption holds only when the characteristic frequencies of turbulence and sound waves differ significantly. When this condition is not met, a substantial portion of acoustic energy may be absorbed by turbulence.

Ingard & Singhal (1974) experimentally investigated acoustic attenuation in turbulent gas flows within pipes, attributing the damping to viscous dissipation and thermal conduction near pipe walls. Howe (1984) theoretically analyzed sound attenuation through turbulent regions, comparing the attenuation coefficient with experimental data from Ronneberger & Ahrens (1977). He concluded that this attenuation arises from strain generated by the acoustic field, with strain magnitudes sharply increasing near solid boundaries. Maximum attenuation occurs when acoustic and turbulent time scales are comparable. Peters et al. (1993) measured the damping and reflection coefficients of plane acoustic waves in smooth pipes (gaseous medium), confirming Ronneberger & Ahrens' findings (1977). Weng, Boij & Hanifi (2013, 2015) conducted theoretical and numerical studies on acoustic attenuation in pipe turbulence, validating results against prior experiments (Ronneberger and Ahrens, 1977; Peters *et al.*, 1993).

Furthermore, the vortex structure inherent in turbulence itself can excite acoustic waves. Building upon Lighthill's acoustic analogy, Howe (2003) developed the Vorticity-Sound Theory, which posits that vorticity is the fundamental source of acoustic radiation in isentropic, low-Mach-number flows. Related research has also extended into the experimental domain (Billon *et al.*, 2008; Wang *et al.*, 2018), where acoustic waves and vortices achieve energy exchange through resonance at similar frequencies.

From the aforementioned studies, it can be observed that experimental research on the interaction between turbulence and acoustic waves has historically focused primarily on aeroacoustics, with limited exploration in hydroacoustics. To our knowledge, the only relevant experiments are those by Ronneberger and Ahrens (1977) on pipe turbulence and Korman and Beyer (1980,1988) on submerged jet turbulence. The observed phenomena of acoustic wave absorption and scattering by turbulence largely align with classical theories. However, these experiments are outdated and limited in scope, failing to fully reveal the underlying mechanisms of hydroacoustic wave propagation in turbulent flows.

In this study, transducers are employed to analyze the characteristics of hydroacoustic waves after interaction with turbulence. Our results demonstrate that acoustic waves can be attenuated or amplified by turbulence, with no observed spectral broadening. These findings, particularly within the tested parameter ranges, cannot be fully explained by conventional theories, providing a new direction for exploring the mechanisms of turbulence.

This paper is organized as follows. First, we present the experimental setup, which includes two configurations: a pipe flow and a submerged water jet. Subsequently, the experimental results obtained under various flow conditions are displayed. Following this, a comparative analysis is conducted, systematically examining potential explanations for the observed signal variations within the framework of classical theories before rejecting them one by one. Finally, we summarize the key conclusions of this study and discuss the follow-up research plan.

## II. EXPERIMENTAL SETUP

The submerged experimental setup comprises two primary configurations: (1) pipe flow (Fig.1), which includes both parallel (sub-divided into co-directional and counter-directional cases) and perpendicular propagation relative to the mean flow direction, and (2) free jet flow (Fig.2). This systematic design allows for comprehensive investigation of flow-acoustic interaction mechanisms, particularly in examining direction-dependent wave modulation effects, where the parallel/perpendicular configurations effectively isolate axial versus transverse interaction components.

In Fig.1(a), two hydroacoustic transducers mounted at opposite ends of the pipeline are connected to a signal generator and an oscilloscope, serving as the acoustic emitter and receiver, respectively. During experiments, the signal generator emits single-frequency sinusoidal waves. External water flow enters through the pipeline inlet and exits at the outlet. Acoustic waves propagating axially along the steel pipe are subject to turbulence modulation. The signal generator controls the incident acoustic wave, while the oscilloscope measures received signals under both static and flowing conditions, including amplitude variations across frequencies.

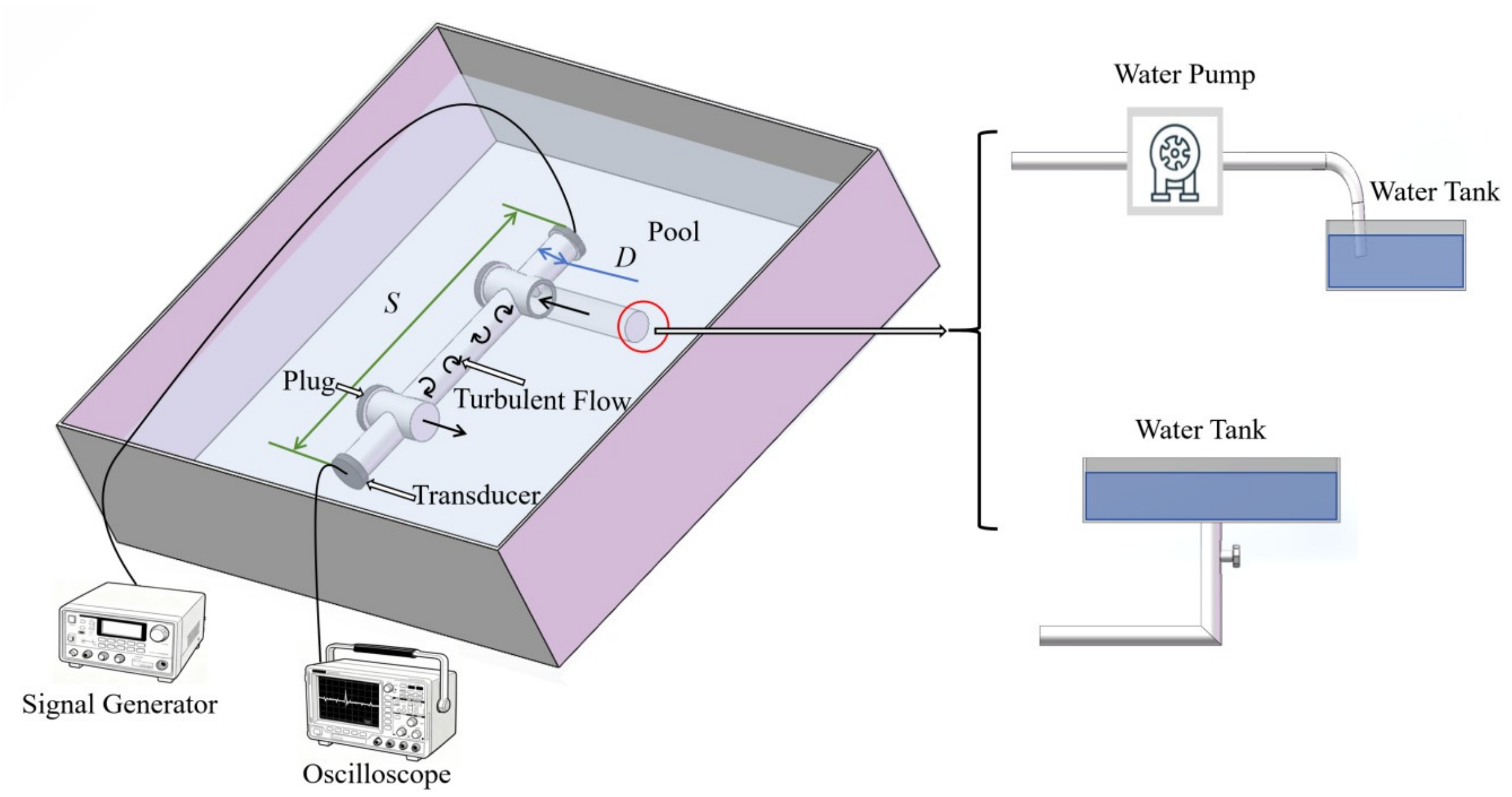

(a)

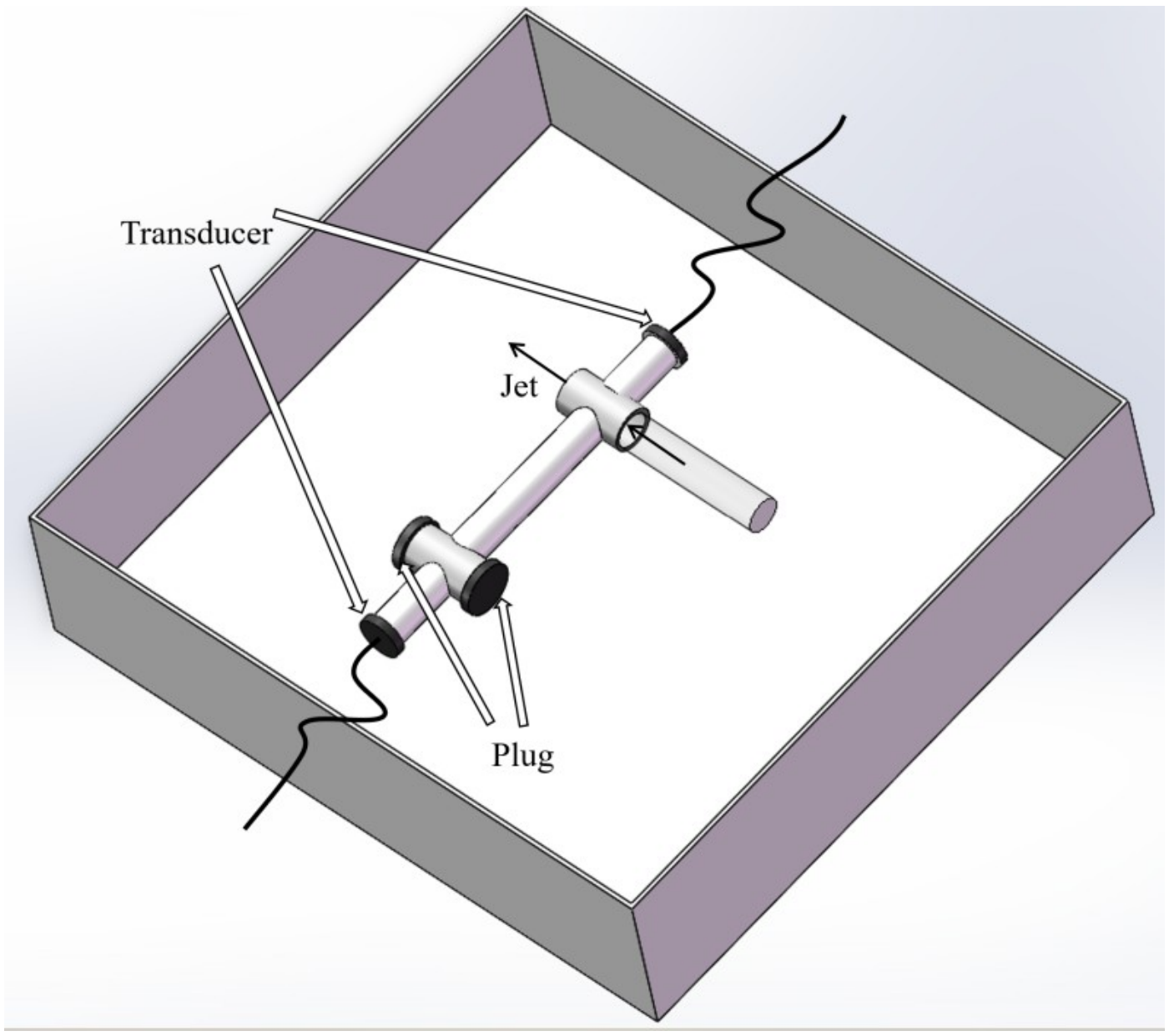

(b)

FIG.1. Schematic diagram of the experimental setup for acoustic wave-turbulence

interaction in pipe flow, showing the relative directions of acoustic propagation and mean flow: (a) parallel; (b) perpendicular.

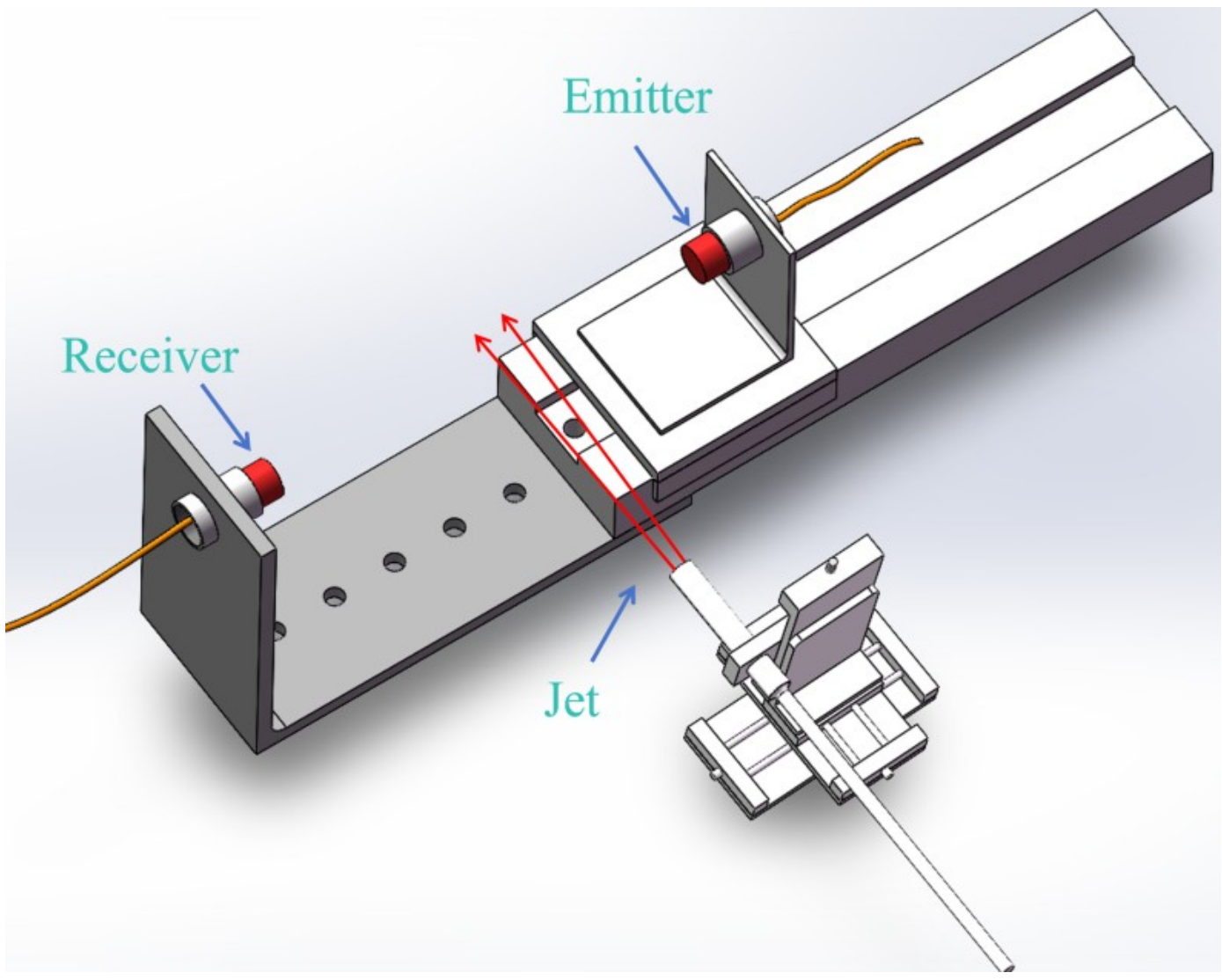

(a)

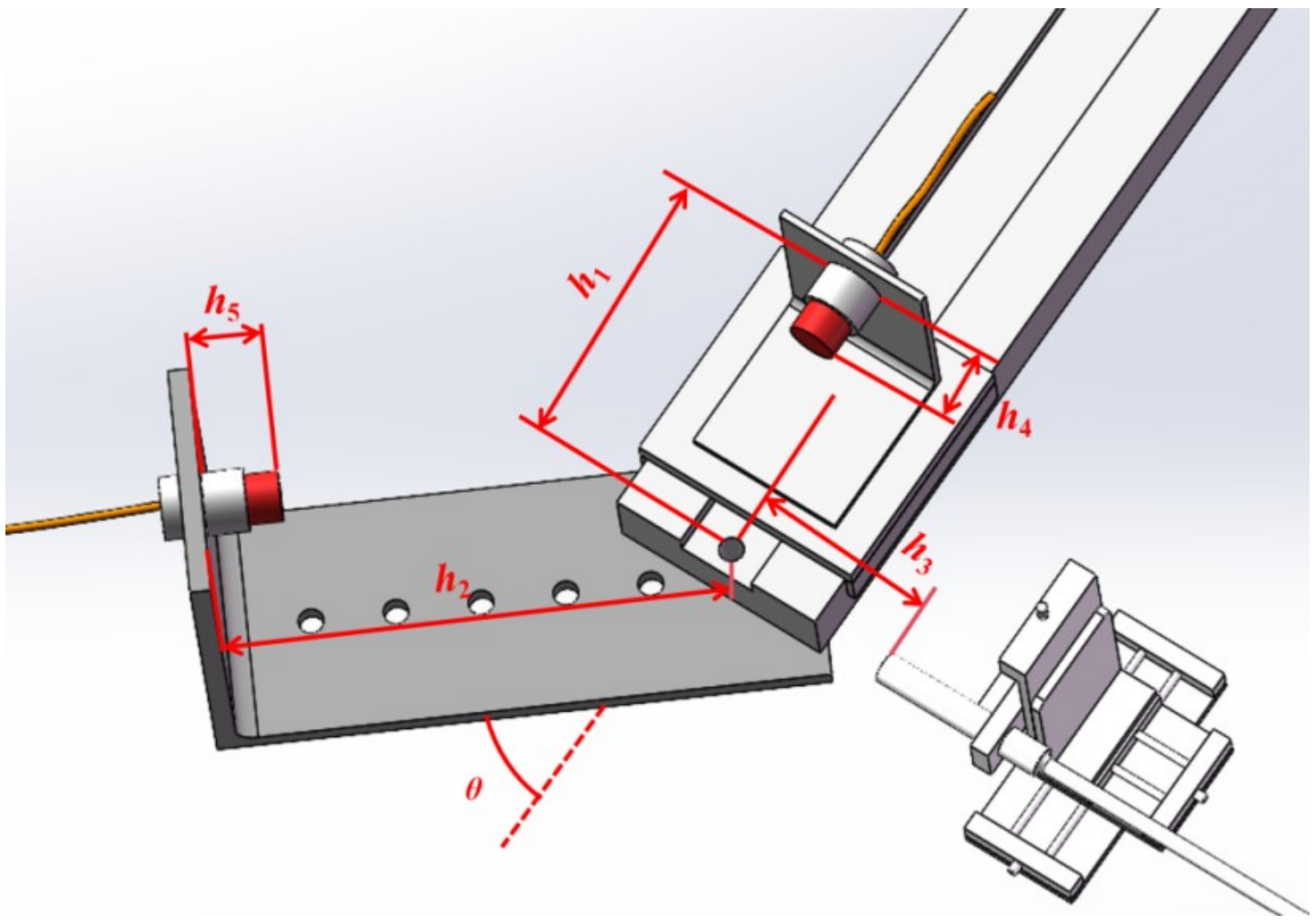

(b)

FIG. 2. Schematic diagram of the experimental setup for acoustic wave-turbulence interaction in a free jet flow, showing configurations of : (a)collinear alignment – emitter and receiver axes coincide vertically with the mean flow direction; (b) angular offset – receiver axis inclined at the angle $\theta$ relative to the emitter.

For the free jet configuration (Fig.2a), the water flow exits through the nozzle into the pool, forming a free jet. Acoustic emitter and receiver are positioned on opposite sides of the jet, with the acoustic propagation direction perpendicular to the mean flow direction. In Fig.2(b), the receiver can be rotated to measure received signals at different angular positions. Detailed parameters for both Figs 1, 2 are provided in Appendix Table VIII.

We use two transducers with a fundamental frequency of 1 MHz as the transmitter and receiver, respectively, and emit acoustic waves near 1 MHz in the experiment. The physical diagram and dimension diagram of the transducers are shown in Appendix. They can emit acoustic waves in the form of a circular piston array, with a half-power angle of 6.2° and a sharpness angle of 14.2°. Thus, most of the energy propagates along the axial direction.

The water flow is driven by two methods: a high-pressure pump and hydraulic head difference (HHD). For the pump configuration, a flowmeter is installed in the inlet pipeline, with a stabilized flow rate of (7.2±0.2)L/min during operation. For the HHD case (Fig.3b), the height difference between the water tank surface and the pool surface is maintained at (1.85±0.05)m during experiments.

## III. RESULTS

### A.Pipe flow

First, we examine the flow state, then discuss the cases when acoustic waves propagate either parallel or perpendicular to the mean flow direction, respectively.

**1. *Examine flow state***

To examine flow states, dye is injected into the pipe. Fig.3 compares three scenarios: (1) flow driven by HHD, (2) pump-driven jet flow, and (3) pump suction flow. In the first two cases, the dye rapidly filled the entire pipe, confirming turbulent flow due to valve-induced disturbances and prolonged wall friction in long pipelines.

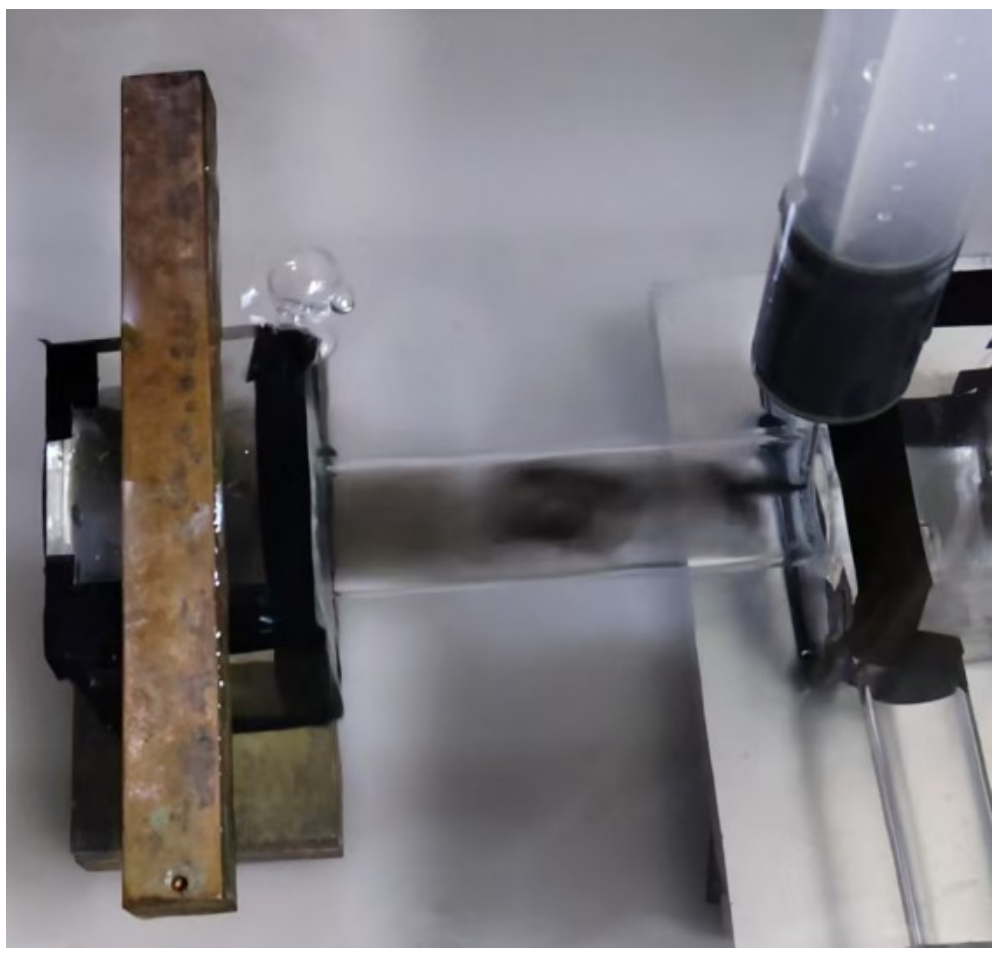

(a)

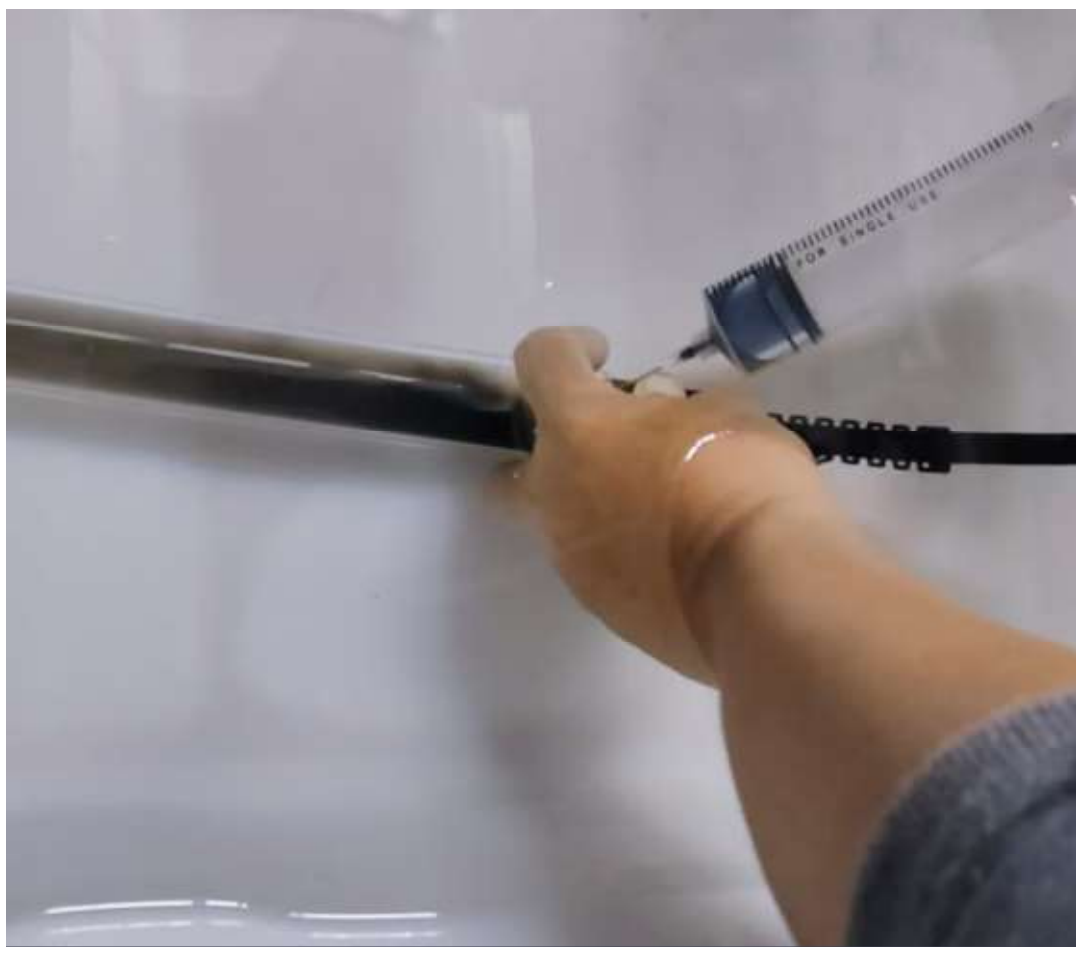

(b)

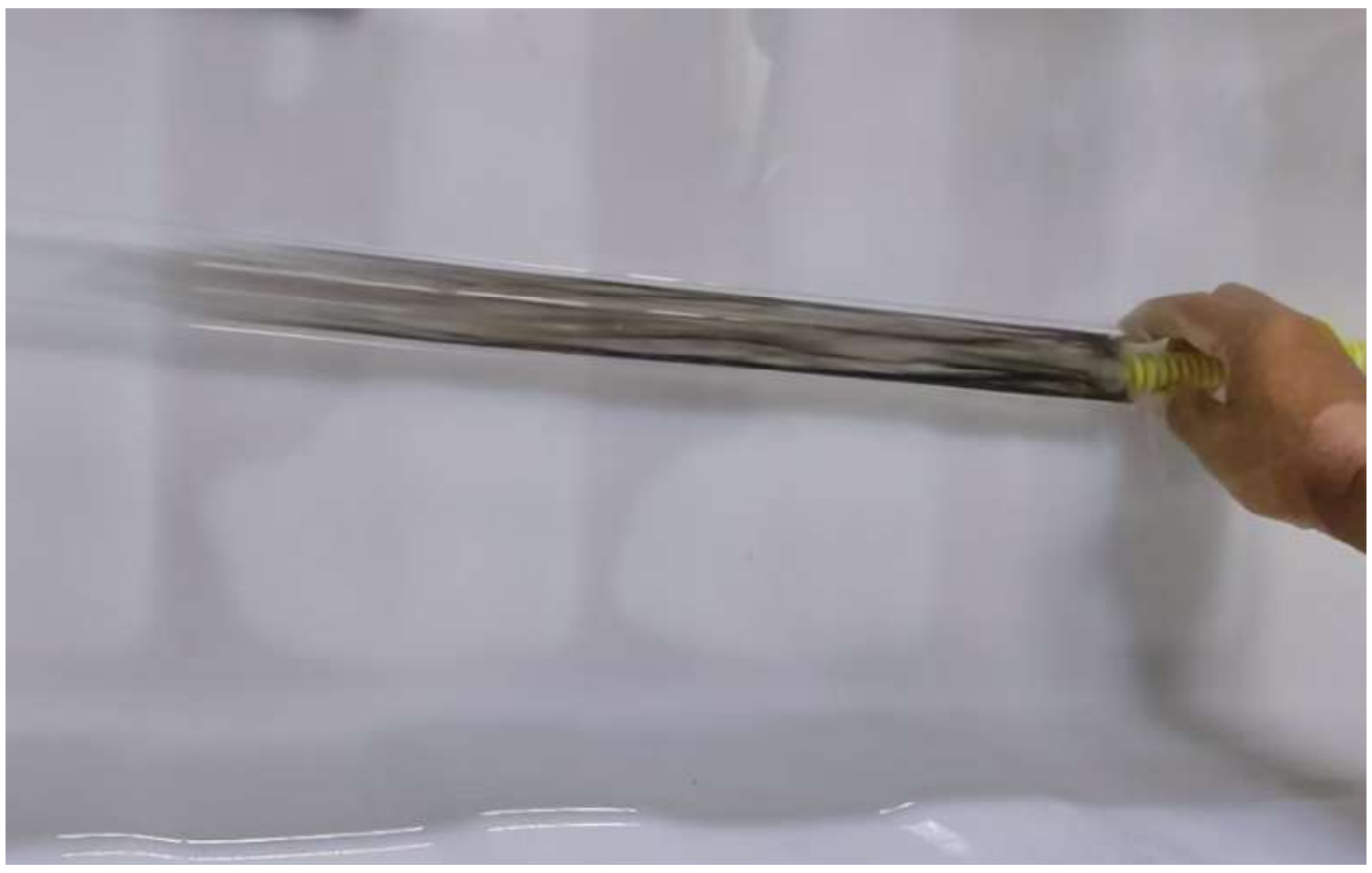

(c)

FIG. 3. The dye visualization reveals distinct flow regimes in the pipe: (a) flow driven by hydraulic head difference; (b) pump-driven jet flow and (c) pump suction flow.

The relevant Reynolds number is estimated as $Re \approx 5094$ , while the energy dissipation rate, the Kolmogorov scale, the friction velocity and the turbulent fluctuation frequency (0-50 Hz range) are determined in Sec.IV B. The turbulence Mach number has the order of $10^{-5}$, which is shown in Sec. IV D. In contrast, the third case exhibited clear, stable dye streamlines, indicating laminar flow—a result of a lower flow rate inside the pipe.

Table I shows the parameters of acoustic waves in pipe flow under various conditions. Here, $V_1$ is the transmitted signal voltage, while $V_2$, and $V_3$ represent the received signal voltage under static water and turbulent flow, respectively. The amplification factor is $A_f = V_3/V_2$ and the relative changes is $V_a = |V_3/V_2 - 1|$.

Table I. Acoustic signals in pipe flow under various conditions, where the error bars represent standard deviation. Tran. Freq. and HHD stand for Transducer's fundamental frequency and hydraulic head difference, respectively.

| No. | Drive Mode | Directions of Acoustic wave and mean flow | Water Temperature (°C) | Emitter | | Receiver | | |
|---|---|---|---|---|---|---|---|---|
| | | | | Frequency $f$(MHz) | Voltage $V_1$(V) | Static $V_2$(mV) | Turbulence $V_3$(mV) | The relative change $V_a$ |
| 1 | HHD | Same | 24.2 | 0.95 | 0.4 | 21.8±0.1 | 17.2±0.1 | ↓,21.1% |
| 2 | HHD | Same | 24.1 | 1.11 | 0.5 | 27.6±0.1 | 40.8±0.2 | ↑,47.8% |
| 3 | Pump | Same | 23.8 | 1.10 | 0.5 | 37.2±0.2 | 25.4±0.6 | ↓,31.9% |
| 4 | Pump | Same | 22.5 | 1.13 | 0.5 | 20.8±0.1 | 24.6±0.2 | ↑,18.1% |
| 5 | HHD | Perpendicular | 22.1 | 0.93 | 0.6 | 18.4±0.1 | 23.6±0.4 | ↑,28.1% |
| 6 | HHD | Perpendicular | 24.1 | 1.03 | 0.3 | 27.2±0.2 | 23.5±0.3 | ↓,13.6% |

**2. *Parallel***

First, we conduct pipe flow experiments with flow driven by HHD. Under static conditions (No. 1 in Table I), the amplitude of the received signal is $V_2$=21.8mv. The acoustic wave propagates in the same direction as the mean flow. When turbulent flow is initiated, the amplitude of the received signal $V_r$ increases, and starts to oscillate after rising to a certain extent. Its average amplitude (arithmetic mean) decreases by 21.1% to $V_3$=17.2mv. When emission is stopped, $V_r$ is reduced to the noise level, less than 0.04% of $V_2$, proving this attenuation results exclusively from turbulence absorption of incident waves, while turbulence alone cannot generate acoustic waves at this frequency. When emission is recovered, $V_r$ returns to its average. After the valve is closed, the received signal gradually returns to the initial stationary value.

Fig. 4 displays the spectra of the acoustic wave altered by turbulence in a pipe flow. Due to the high operating frequency, the received signal spectrum displayed on the oscilloscope exhibits measurable bandwidth. Spectral analysis reveals: (1) consistent peak amplitude at 950.0kHz (vs. 950.1kHz minimum, Fig.4a), (2) synchronous amplitude variations across all frequencies, and (3) no spectral broadening within the oscilloscope's 1Hz resolution - demonstrating perfect frequency preservation between emitted and received signals. These findings validate that turbulent flow under these conditions attenuates incident acoustic waves without inducing frequency shifts.

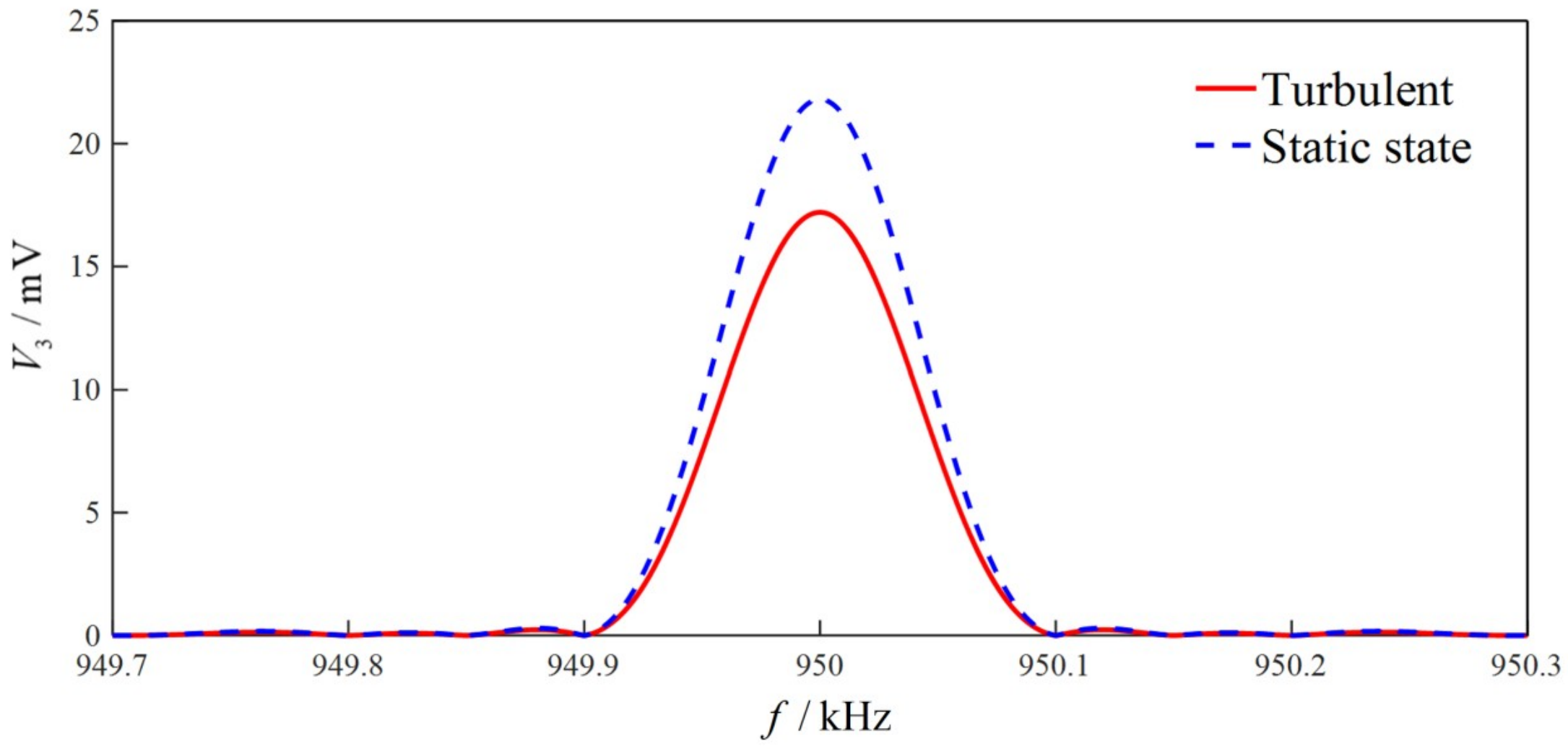

(a)

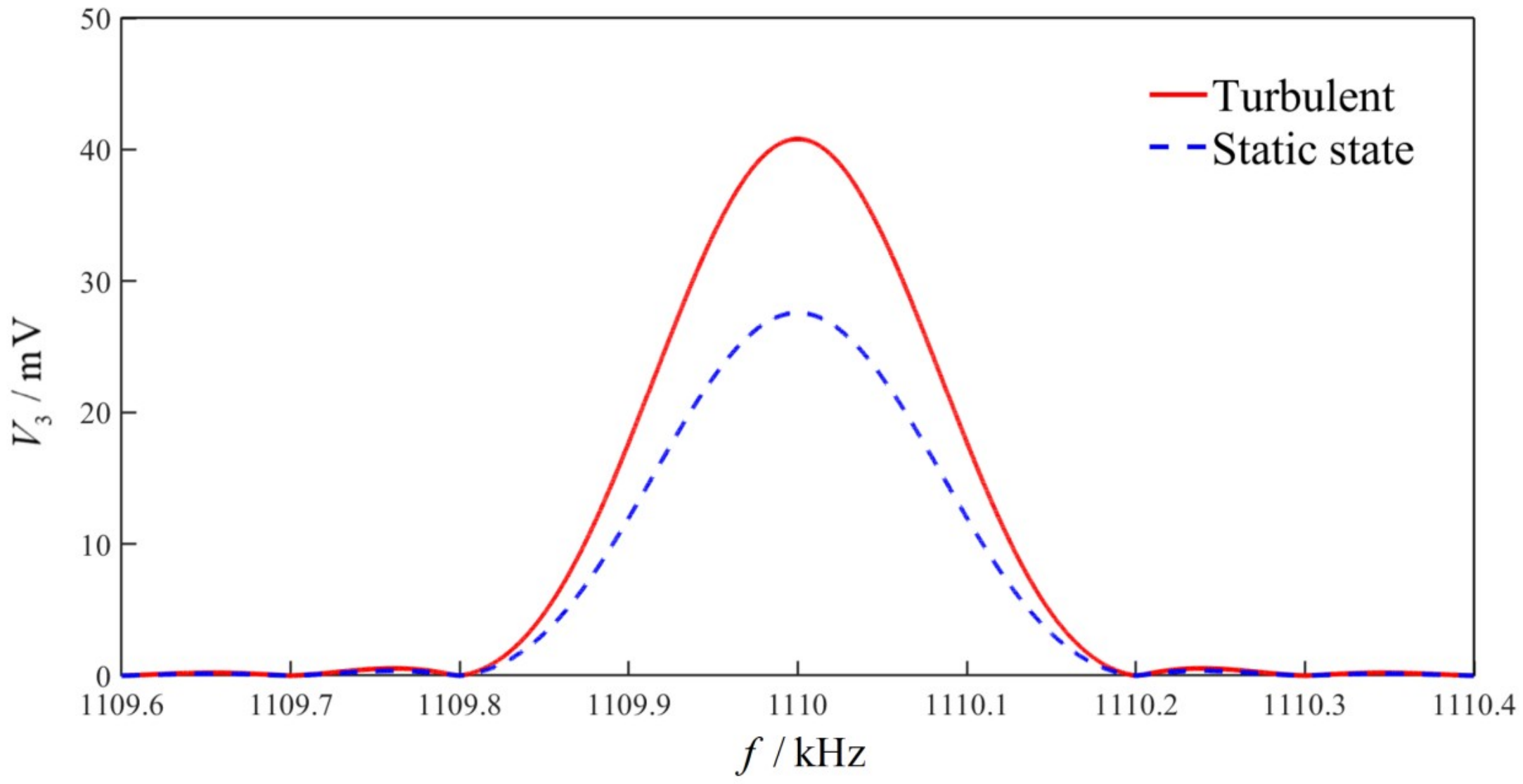

(b)

FIG. 4. Spectra of the acoustic wave altered by turbulence in a pipe flow: (a) 0.95MHz; (b) 1.11MHz. The turbulent flow is driven by the HHD.

When the acoustic emitter emits a 1.11 MHz signal at 0.5V (No. 2 in Table I), the receiver amplitude increases from $V_2$=27.6 mV (static water) to an average $V_3$=40.8 mV (turbulent flow), representing a 27.7% increase. Signal termination tests confirmed immediate amplitude drop to the noise level upon emitter deactivation, confirming that the incident wave is amplified by the turbulence, while turbulence alone is not capable of generating acoustic waves at this frequency. Spectral analysis (Fig.4b) also shows no spectral broadening. Combined with prior 0.95 MHz experiments, these results conclusively demonstrate frequency-dependent acoustic modulation in pipe flows.

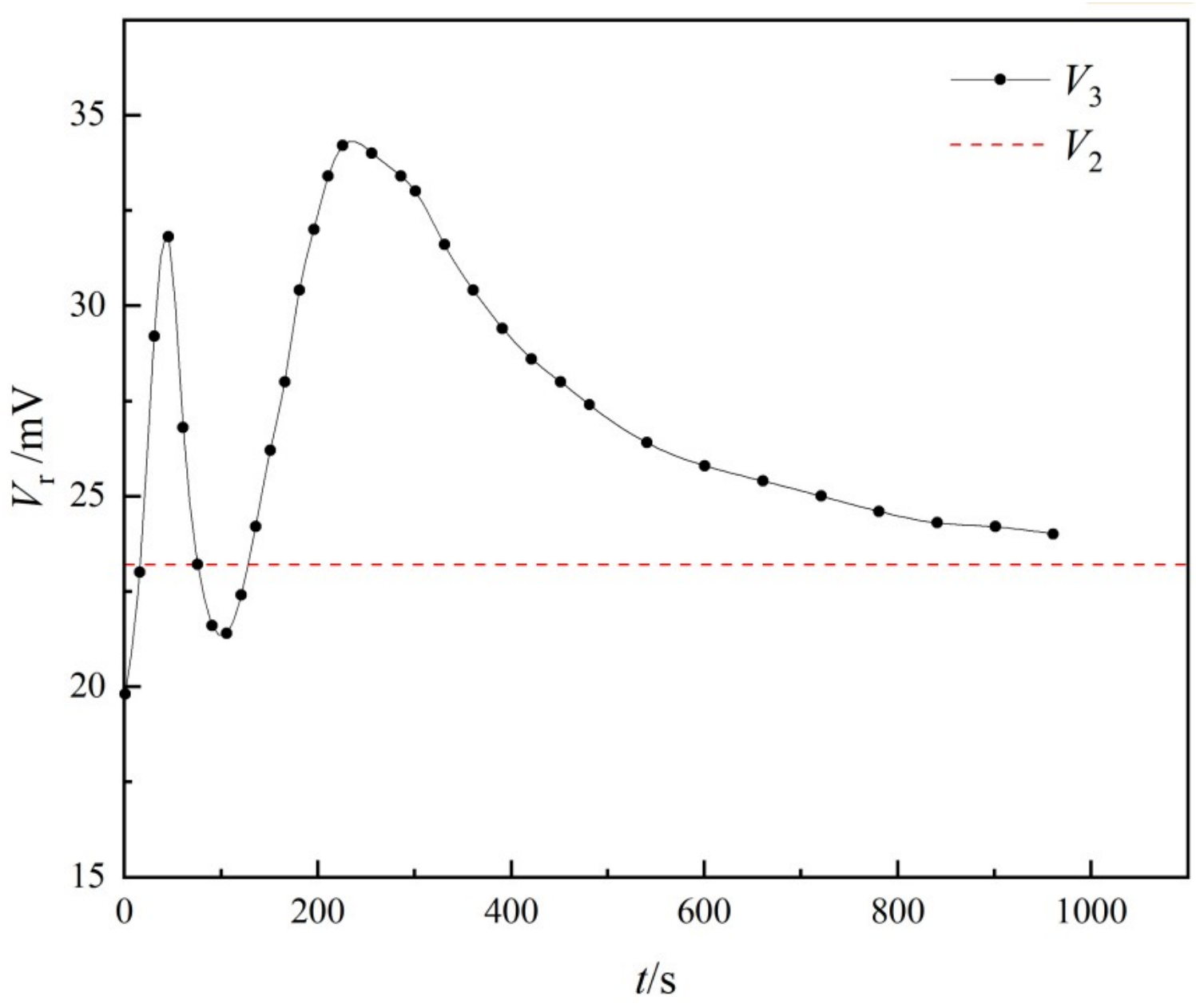

FIG. 5. The variation of amplitude with time for the acoustic wave altered by turbulence in a pipe flow after the valve is closed. The emission is 1.10 MHz at 0.4V, and the water temperature is 23°C.

Our experimental observations reveal a significant delay in signal recovery after valve closure, with the receiver taking an extended period to stabilize at the original static value $V_2$. Fig.5 demonstrates the variation of acoustic amplitude with time in a pipe flow after the valve is closed. At the beginning, the turbulent flow is driven by the HHD. When the valve is closed ($t$=0s), $V_3$ =19.8mV; after two significant oscillations, it gradually approaches the static value $V_2$=23.2mV. The maximum drop during this period is 14.7%, and the maximum rise is 47.4%.

This phenomenon occurs because while the valve closure immediately reduces mean flow velocity to zero, turbulent fluctuations within the pipe persist due to fluid inertia. These residual turbulent motions gradually dissipate through viscous damping, during which the receiver signal oscillates before ultimately returning to its baseline level. These results also indicate that turbulent fluctuations of different intensities lead to different degrees of acoustic amplification, and that both amplification and attenuation can occur during the decay of turbulence.

To further investigate this phenomenon, the turbulent flow is injected into the pipe inlet using a pump nozzle (Fig. 6a, No. 3 in Table I). The frequency of acoustic wave is 1.1MHz. When the received signal changes ($V_3$= 25.4 mV), the nozzle is removed. Notably, the signal does not immediately recover to its static value ($V_2$=37.2 mV).

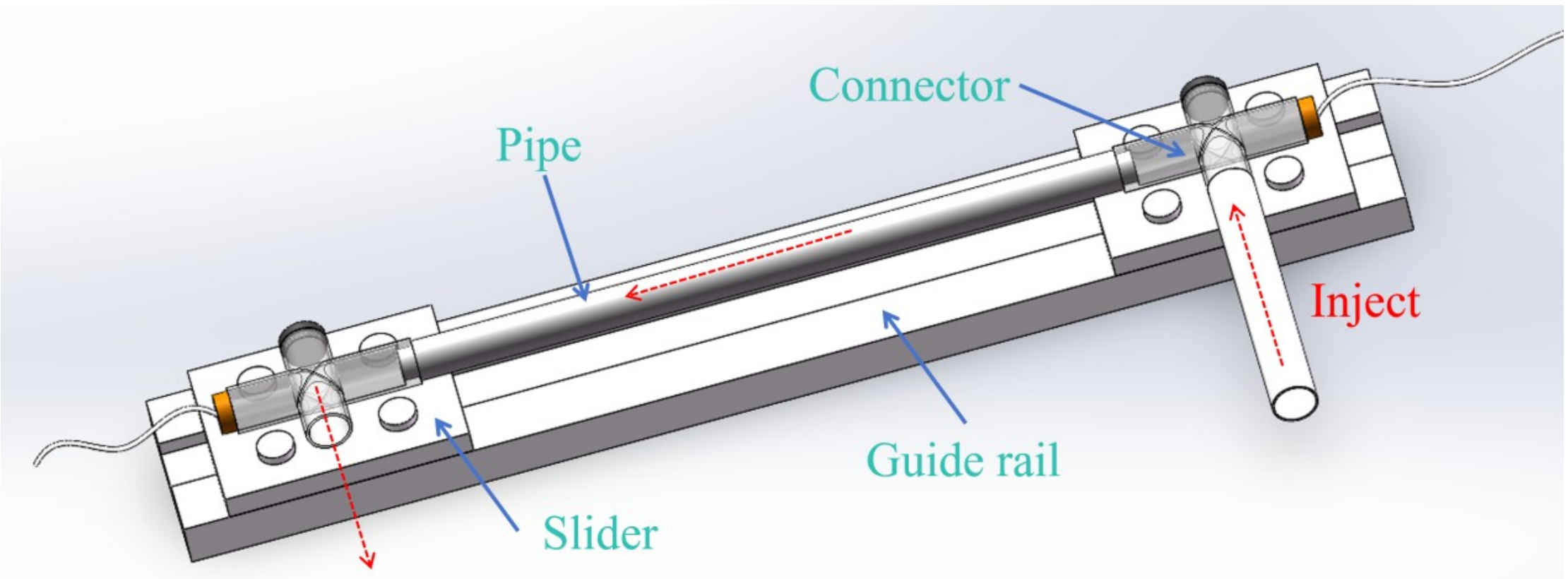

(a)

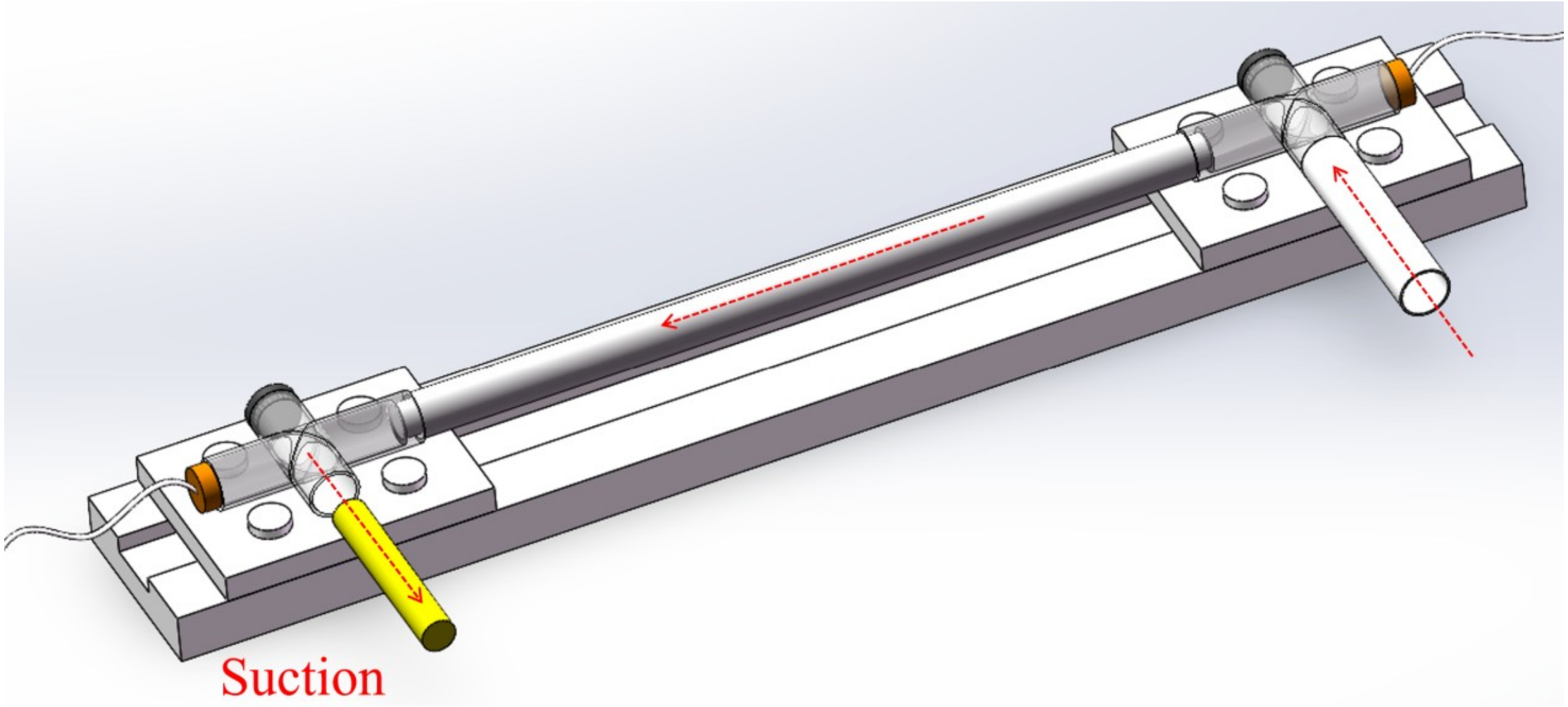

(b)

FIG. 6. Experimental setup of acoustic-turbulence interaction in a pipe flow driven by a pump : (a) Inject at the inlet; (b) Suction at the outlet.

Subsequently, we apply suction near the pipe outlet using the pump's intake (Fig.6b). This intervention causes the received signal to rapidly return to $V_2$. The restoration occurs because suction changes the internal flow into laminar regime, eliminating turbulent fluctuations and thus recovering the acoustic amplitude. The differential recovery behavior under identical flow parameters proves that: acoustic amplitude variation depends on turbulence intensity; laminar flow conditions permit complete signal restoration; mean flow alone cannot explain observed signal changes.

For the wave with a frequency of 1.13 MHz, input voltage $V_1$ = 15V (No. 4 in Table I), the received wave amplitude decreases from $V_2$ = 20.8 mV (static water) to $V_3$ = 24.6mV under turbulent flow—an increase of 18.1%. In the pump-driven flow, the variation of the spectrum is similar to that in Fig. 6, and there is likewise no spectral broadening.

If the positions of the inlet and outlet of the pipeline are swapped in Fig.6a, acoustic waves can propagate against the direction of the mean flow.Table II shows a comparison of the received signal amplitudes when the acoustic waves travel in the same direction as and opposite to the mean flow. It can be observed that the $V_3$ values for both cases are very close, with a relative deviation within 1%. Therefore, it can be concluded that the direction of the mean flow has no substantial impact on the acoustic waves.

Table II Acoustic signals in pipe flow when the acoustic waves travel in the same direction as and opposite to the mean flow.

| No. | Drive Mode | Directions of Acoustic wave and mean flow | Tran. Freq. (MHz) | Water Temperature (°C) | Emitter | | Receiver | | |
|---|---|---|---|---|---|---|---|---|---|
| | | | | | Frequency (MHz) | Voltage $V_1$(V) | Static $V_2$(mV) | Turbulence $V_3$(mV) | The relative change $V_a$ |
| 1 | Pump | Same | 1 | 23.8 | 1.10 | 0.5 | 37.2±0.2 | 25.4±0.6 | ↓，31.9% |
| 2 | Pump | Opposite | 1 | 23.8 | 1.10 | 0.5 | 37.2±0.2 | 25.5±0.9 | ↓，31.6% |
| 3 | Pump | Same | 1 | 22.5 | 1.13 | 0.5 | 20.8±0.1 | 24.57±0.2 | ↑，18.1% |
| 4 | Pump | Opposite | 1 | 22.5 | 1.13 | 0.5 | 20.8±0.1 | 24.54±0.2 | ↑，18.0% |

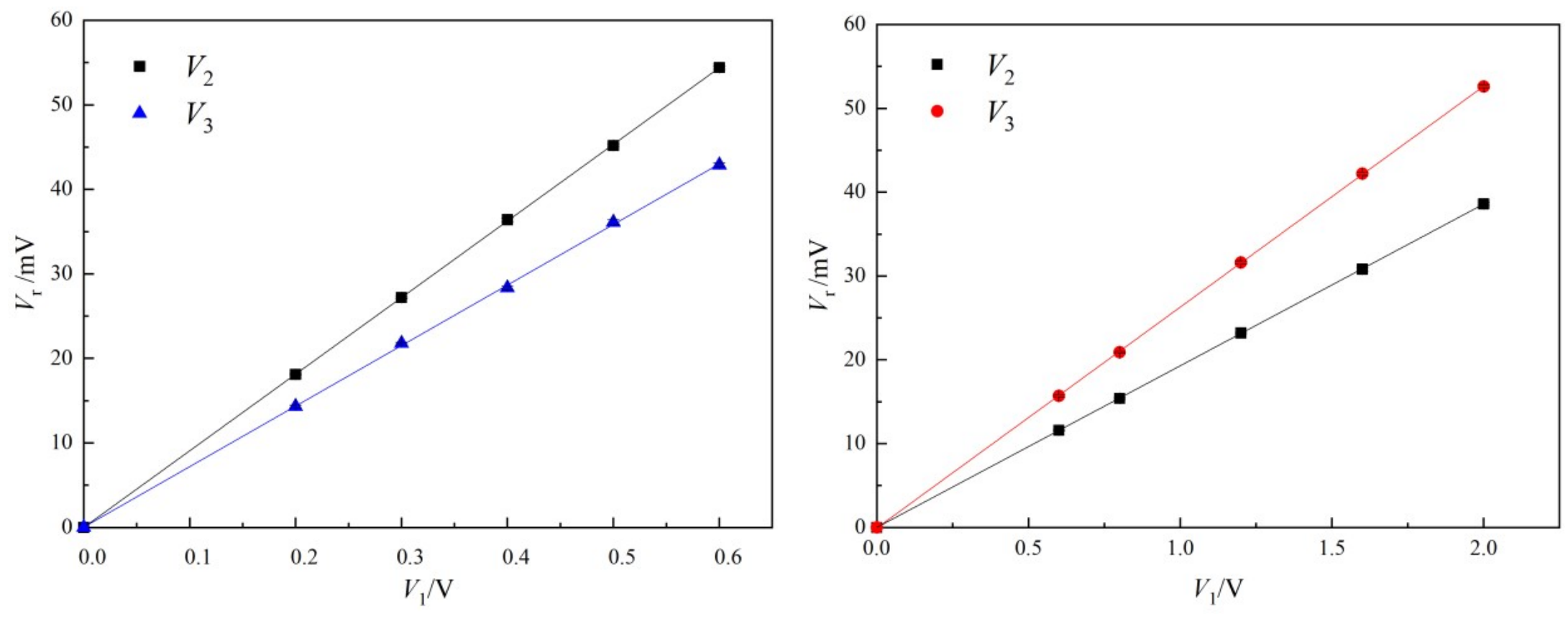

(a) (b)

FIG. 7. The voltage values of the received signal $V_r$ under different voltages of the transmitted signal $V_1$. The frequency of the transmitted signal $f$ is (a)1.0 MHz and (b)1.2 MHz, respectively. The water temperature is 23.3°C, turbulence is injected by a water pump, and the acoustic wave propagates in the same direction as the mean flow.

Fig.7 displays the voltage values of the received signal $V_r$ under different voltages of the transmitted signal $V_1$. Here, $V_r$ includes $V_2$ and $V_3$. The standard deviation in the figure is very small, so most error bars cannot be seen. It can be seen that $V_2$ and $V_3$ are all approximately proportional to $V_1$. This linear proportional relationship holds true for both the attenuation case (Fig.7a) and the amplification case (Fig.7b). Therefore, the amplification factor $A_f$ is independent of the input voltage $V_1$. For $f$=1.0 MHz, $A_f$=0.78 , and for $f$=1.2 MHz, $A_f$=1.36.

### 3. *Perpendicular*

When the acoustic waves propagate perpendicular to the mean flow, experiments are conducted with different frequencies and driving methods (Fig.1(b), Nos. 5,6 in Table I). The results also show that turbulence can change the wave amplitude with no spectral broadening. Suction near the pipe opening (Fig.8) does not alter the received signal.

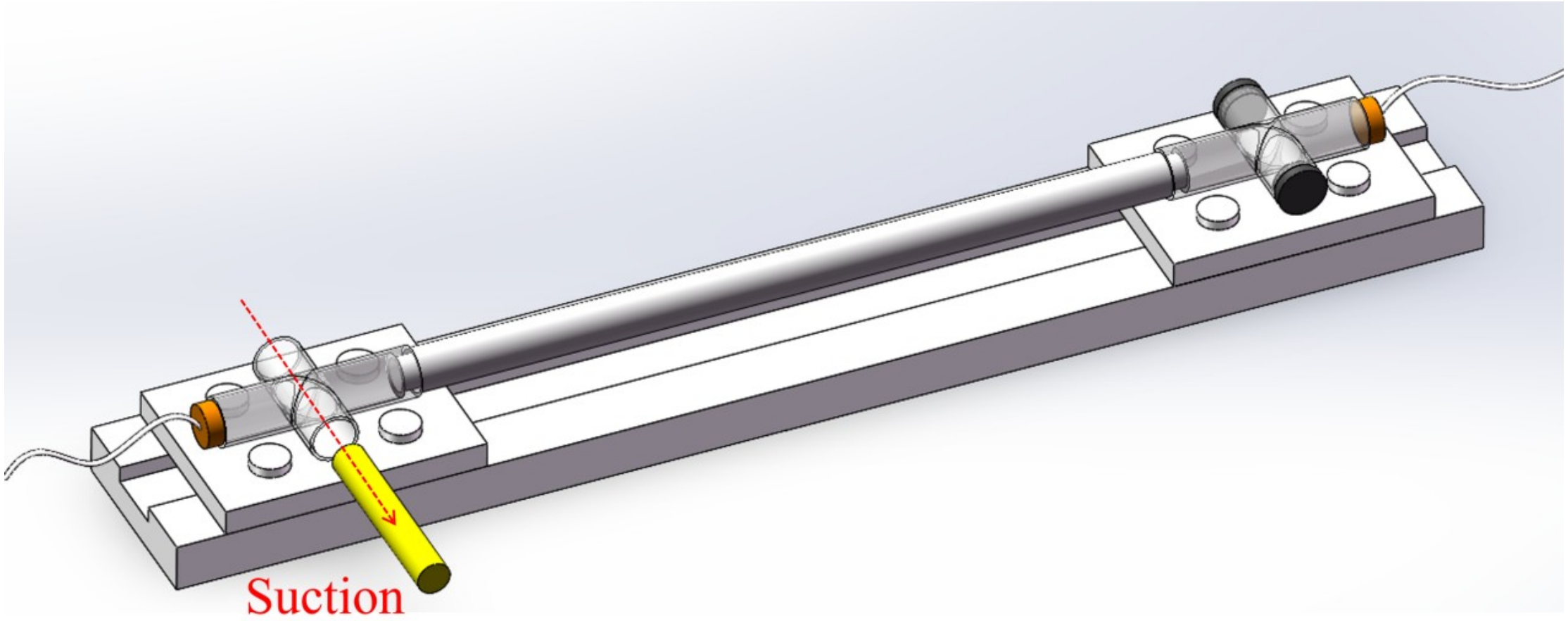

FIG.8. Experimental setup for acoustic waves propagate perpendicular to the mean flow: suction at the outlet driven by a pump.

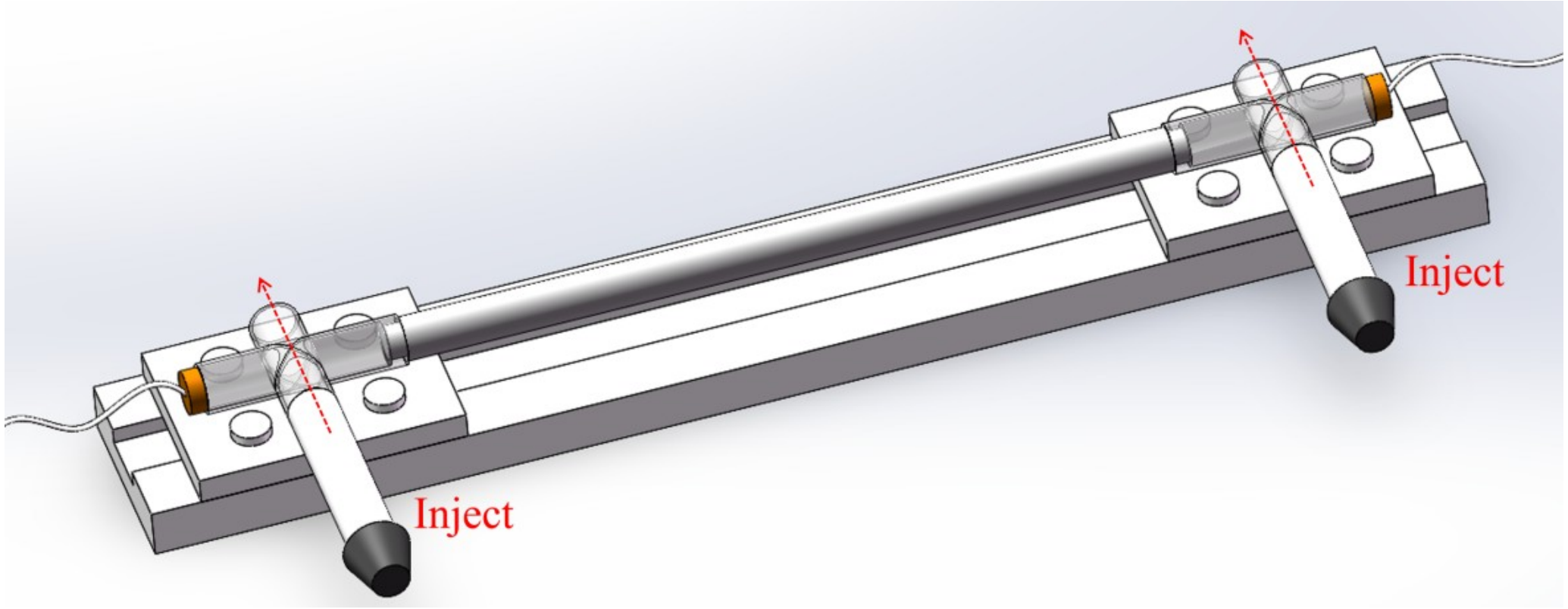

FIG. 9. Experimental setup of acoustic waves propagate perpendicular to the mean flow with two water inlets. Here, the two inlets are supplied with water flow by two water pumps, respectively.

Based on the properties of the amplification factor, we hypothesize that the total effect of attenuation (amplification) in the entire pipeline should be equal to the product of the effects in each segment. To verify this hypothesis, the following experiment is conducted (Fig.9): two water inlets are installed in the pipeline, and water flow is injected through each inlet by two separate water pumps. This setup creates two regions with concentrated turbulence in the pipeline, and the two regions are approximately independent of each other.

Table III shows the received signal values when the two water pumps are activated separately and jointly. Among them, $V_{r1}$($V_{r2}$) represents the amplitude of the received signal when a single water pump is activated to inject water into the inlet close to the transmitting end (receiving end). $V_{r12}$ denotes the amplitude of the received signal when both water pumps are activated simultaneously. The amplification factors of the acoustic signal when the two water pumps are independently activated are $V_{r1}/V_2, V_{r2}/V_2$, respectively. The theoretical total amplification factor when both water pumps are activated simultaneously is the product of the amplification factors when they are independently activated, i.e., $(V_{r1} \cdot V_{r2})/V_2^2$, while the actual measured value is $V_{r12}/V_2$. It can be observed that the theoretical value is very close to the actual measured value.

Table III. Amplitude of acoustic waves in pipelines with two water inlets under various conditions.

| No. | Temperature (°C) | Emitter | | | Receiver | | | | |
|---|---|---|---|---|---|---|---|---|---|
| | | Frequency (MHz) | Voltage $V_1$(V) | Static $V_2$(mV) | $V_{r1}$(mV) | $V_{r2}$(mV) | $V_{r12}$(mV) | Theoretical total amplification factor $\frac{V_{r1} \cdot V_{r2}}{V_2^2}$ | Actual total amplification factor $\frac{V_{r12}}{V_2}$ |
| 1 | 16.1 | 0.90 | 1.5 | 12.5±0.1 | 9.84±0.06 | 10.0±0.04 | 7.84±0.04 | 63.0% | 62.7% |
| 2 | 16.5 | 1.00 | 0.5 | 16.2±0.1 | 17.4±0.1 | 17.8±0.1 | 19.2±0.1 | 118.0% | 118.5% |
| 3 | 15.8 | 1.10 | 0.2 | 10.04±0.04 | 9.32±0.04 | 9.56±0.04 | 8.76±0.04 | 88.4% | 87.3% |
| 4 | 16.5 | 1.20 | 1.5 | 14.6±0.1 | 12.7±0.1 | 12.3±0.1 | 10.76±0.04 | 73.3% | 73.7% |

**B. Free Jet**

For the free jet configuration, the emitter and receiver are positioned on opposite sides of the jet (Fig.2), with the receiver being rotatable. Based on the flow rate $Q_2$ ≈7.0 L/min and pipe diameter $D_2$=1.7cm, the Reynolds number of the flow is approximately $Re$≈9000.

First, we examine the case where the azimuth angle $\theta$=0° (acoustic waves propagate perpendicular to the mean flow). For acoustic waves at a frequency of approximately 1 MHz, the wavelength has $\lambda \approx 1.5\text{mm}$ , the piston radius is $a$ = 0.85 cm, and the Fresnel distance has $a^2/\lambda \approx 4.82\text{cm}$ . The distance between the end faces of the two transducers is $L_e$≈21 cm. As $L_e >> a$ and $L_e >> a^2/\lambda$ , the far-field condition is satisfied. For the cases 1~4 in Table IV, with jet turbulence present, the received signals show distinct increases and decreases respectively compared to static state. The amplitude variations are smaller than those observed in pipe flow conditions. No Doppler shift or spectral broadening are detected.

The received signals at various azimuth angles are measured (Table IV). The experimental data reveal signal attenuation: rapid signal degradation occurs when $\theta$≥15°. Negligible amplitude variation is observed before/after jet initiation. Similar attenuation patterns are observed for $\theta$<0°. These findings confirms the absence of significant turbulent scattering to other angular directions.

Table IV. Acoustic signals in free jet under various conditions.

| No. | Drive | azimuth | Water | Emitter | Receiver |
|---|---|---|---|---|---|

| | Mode | angle $\theta$ (°) | Temperature (°C) | Frequency (MHz) | Voltage $V_1$(V) | Static $V_2$(mV) | Turbulence $V_3$(mV) | The relative change $V_a$ |
|---|---|---|---|---|---|---|---|---|
| 1 | HHD | 0 | 21.0 | 0.943 | 0.8 | 21.4±0.1 | 23.5±0.2 | ↑,9.8% |
| 2 | HHD | 0 | 22.1 | 0.983 | 0.5 | 28.4±0.1 | 26.0±0.2 | ↓,8.5% |
| 3 | Pump | 0 | 28.8 | 1.237 | 1.2 | 17.8±0.1 | 19.7±0.4 | ↑,10.7% |
| 4 | Pump | 0 | 28.8 | 0.966 | 0.8 | 29.6±0.1 | 25.1±0.9 | ↓,15.2% |
| 5 | Pump | 15 | 28.8 | 0.966 | 0.8 | 1.24±0.04 | 1.24±0.08 | — |
| 6 | Pump | 30 | 28.8 | 0.966 | 0.8 | 0.36±0.04 | 0.38±0.07 | — |
| 7 | Pump | 60 | 28.8 | 0.966 | 0.8 | 0.44±0.04 | 0.46±0.08 | — |

## IV. ANALYSIS

Below, we attempt to identify the mechanisms of acoustic amplitude modulation by turbulence from conventional theories, systematically eliminating potential explanations while conducting comparative analysis with previous literature.

### A. Bubble

The bubbles inside the water can scatter acoustic wave, leading to significant attenuation of the incident sound waves (Leighton, 1994; Richards and Leighton, 2003; Zheng and Huang, 2016). In addition, scattering caused by bubbles will introduce new frequency spectra (De Jong *et al.*, 2002; Goertz *et al.*, 2003; Yoshida *et al.*, 2008). However, in our experiments, the amplitude of the acoustic wave can increase under the influence of turbulence, while the frequency spectrum remains unchanged. These phenomena cannot be explained by bubbles.

In a jet flow, if the velocity at the outlet is too high, cavitation may occur in the liquid, leading to the formation of bubbles. A commonly used dimensionless number in cavitation study is the cavitation number, which is defined as

$$\sigma = \frac{P_\infty - P_\mathrm{v}}{\frac{1}{2}\rho V_\mathrm{c}^2}. \tag{4.1}$$

Here $P_\infty$ is the downstream pressure, $P_\mathrm{v}$ is the vapor pressure of the water, $\rho$ is the density and $V_\mathrm{c}$ is the average liquid velocity at the orifice (Yan and Thorpe 1990). Cavitation only occurs when $\sigma$ falls below a certain critical value (Cavitation Inception Number $\sigma_\mathrm{i}$). In engineering, $\sigma_\mathrm{i}$ typically ranges from approximately 1.7 to 2.4 (Yan and Thorpe 1990).

In our experiment, we choose the parameters at 18°C: the saturated vapor pressure of water is $P_\mathrm{v} = 2.06\mathrm{KPa}$, the density is $\rho \approx 10^3 \mathrm{kg/m^3}$, and the standard atmospheric pressure $P_\infty = 101.33\mathrm{KPa}$. By connecting a flowmeter to the pump, the measured flow rate during operation is approximately $Q$ = 7.2 L/min. When the nozzle diameter is $D_0$=7mm, the mean velocity from the nozzle can be calculated using the continuity equation:

$$v_1 = \frac{4Q}{\pi D_0^2} \approx 3.12\mathrm{m/s}. \tag{4.2}$$

Thus, the cavitation number has

$$\sigma = \frac{P_\infty - P_\mathrm{v}}{\frac{1}{2}\rho v_1^2} \approx 20.4, \tag{4.3}$$

which is far lager than the range of $\sigma_{\mathrm{i}}$.

For a gravity-driven flow setup (Fig.3b), based on the rate of water level decline, the calculated flow rate is approximately $Q_2$ =7.0 L/min. For the pipe diameter $D_2$ = 1.7 cm shown in Fig.5, the mean velocity inside the pipe is:

$$v_2 = \frac{4Q_2}{\pi D_2^2} \approx 0.51\mathrm{m/s}. \tag{4.4}$$

Its cavitation number is even larger that in (4.3). Therefore, cavitation is impossible in both cases.

In addition, during the experiments, one can observe that after closing the valve at the inlet, only turbulent fluctuations remain in the pipe, and it take a considerable amount of time (ranging from a few minutes to over twenty minutes) for the wave amplitude to return to the level observed in static water (Fig.5). During this period, the mean flow in the pipe ceases, and the turbulent fluctuations gradually decays. Even if we assume that cavitation bubbles exist in the pipe, they would rapidly collapse as the fluid pressure return to the atmospheric level, making it impossible for them to affect the acoustic wave over such an extended duration.

Therefore, the influence of bubbles can be ruled out in these experiments.

**B. Resonance**

Howe (1984) noted that acoustic wave attenuation primarily occurs when the timescales of the sound waves and turbulence are comparable, i.e., a resonance mechanism. To address this, the fluctuation frequencies generated by turbulence are measured and calculated in the following.

First, we measure the frequency spectrum of turbulent fluctuations. The experimental setup is the same as that in Fig.6(a), while the fundamental frequency of the transducer is 7 kHz. It can also detect low-frequency acoustic waves (below 1kHz). Fig.10 shows the receiver spectrum when no acoustic signal is actively transmitted. Without turbulence, only a slight peak of 0.24 mV is observed at 50 Hz, corresponding to the AC power frequency of the oscilloscope. With turbulence , the 50 Hz peak increased to 0.32 mV (likely due to the 50 Hz AC power input driving the water pump). Additional low-frequency peaks emerged below 50 Hz (e.g., a 29.1 Hz peak of 0.48 mV). When the jet nozzle is moved away from the pipe inlet (disrupting turbulence generation), these spectral peaks disappear. So one can confirm that the turbulent spectrum is concentrated within the 0-50 Hz range.

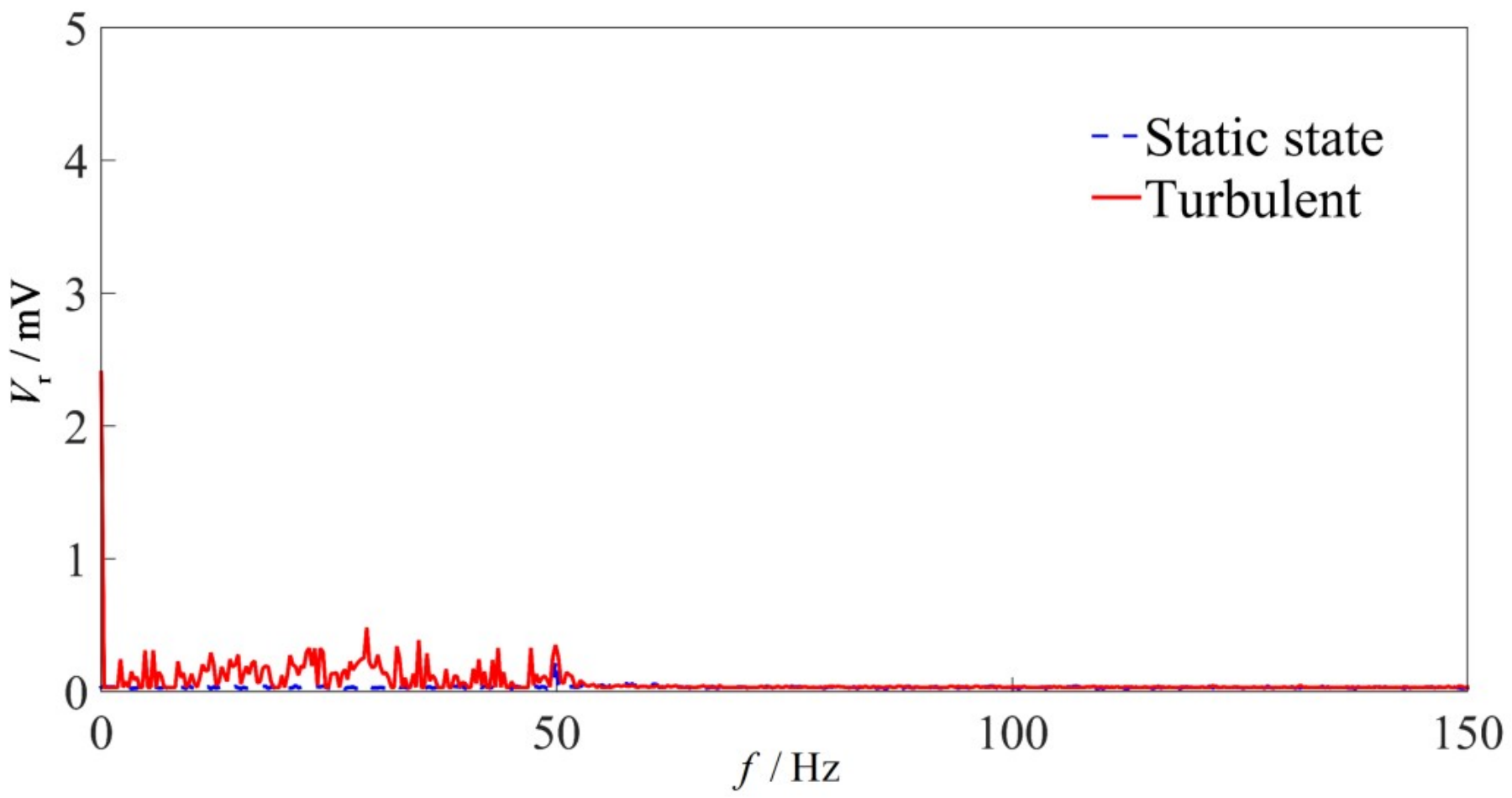

FIG. 10. Receiver spectrum when no acoustic signal is actively transmitted.

The turbulence spectrum can also be estimated theoretically. In Fig.10, the pipe diameter $D_1$=3cm. The characteristic velocity can be estimated as $U_1 = Q/A \approx 16.98\text{cm/s}$, where $A = \frac{1}{4}\pi D_1^2$ is the cross-sectional area of the pipe. At room temperature 18°C, the kinematic viscosity of water $\nu \approx 10^{-6}\text{m}^2/\text{s}$. The Reynolds number of the flow is $Re = U_1 D_1/\nu \approx 5094$. For simplicity, the flow within the pipe is assumed to be fully developed turbulence.

In order to examine the dissipation range characterized by the Kolmogorov scale, the friction velocity $u_*$ is estimated as follows. It can be derived from the mean velocity $U_1$ and the Darcy-Weisbach friction factor $\lambda$. For smooth-pipe turbulence, a widely used approximation is the Blasius formula (see Eq. 20.5 in Schlichting & Gersten (2016)):

$$\lambda \approx 0.3164 \cdot Re^{-0.25}. \tag{4.5}$$

Substituting $Re$ = 5094, it has $\lambda \approx 0.03745$. Then, according to the wall shear stress formula (Eq. 19.24) and (20.4) in Schlichting & Gersten (2016):

$$\tau_1 = \rho(u_*)^2, \tag{4.6}$$

$$\tau_1 = \frac{1}{8}\lambda\rho U_1^2, \tag{4.7}$$

so

$$u_* = U_1\sqrt{\frac{\lambda}{8}} \approx 1.162\text{cm/s}. \tag{4.8}$$

Then, the energy dissipation rate can be calculated. In fully developed pipe turbulence, the energy dissipation rate $\varepsilon$ can be estimated through global energy balance. For a pipe segment of length $L$, the power input by the pressure difference $\Delta P$ balances the turbulent dissipation power:

$$\dot{W} = \Delta P \cdot Q = \varepsilon L A \rho. \tag{4.9}$$

Furthermore, based on force balance,

$$\Delta P \cdot A = \tau_1 \pi D_1 L. \tag{4.10}$$

Thus,

$$\varepsilon = \frac{4\tau_1 U_1}{\rho D_1} = \frac{4U_1 (u_*)^2}{D_1}. \tag{4.11}$$

Substituting all values, it has $\varepsilon \approx 3.056 \times 10^{-3} \mathrm{m}^2/\mathrm{s}^3$.

Later, the Kolmogorov scale $\eta$ cam be estimated. This defines the characteristic size of the smallest eddies, whose energy is dissipated through viscous effects. Its formula is (Pope, 2000):

$$\eta \approx \left(\nu^3/\varepsilon\right)^{\frac{1}{4}}, \tag{4.12}$$

substituting all values, one can obtain $\eta \approx 1.345 \times 10^{-4} \mathrm{m}$. The characteristic velocity, time and frequency are

$$v_\eta \approx (\varepsilon \nu)^{\frac{1}{4}}, \tag{4.13}$$

$$t_\eta \approx \sqrt{\frac{\nu}{\varepsilon}} \quad f_\eta = \frac{1}{t_\eta} \approx \sqrt{\frac{\varepsilon}{\nu}}. \tag{4.14}$$

Substituting all values, one can derive $v_\eta \approx 0.7435 \mathrm{cm/s}$ , $f_\eta \approx 55.28 \mathrm{Hz}$.

For the frequency in the energy-containing range, we take the characteristic length as the pipe radius and the characteristic velocity as the friction velocity or the mean flow velocity, resulting in the large-eddy turnover frequency

$$f_{\mathrm{L1}} = \frac{u_*}{D_1/2}, \tag{4.15}$$

and the flow-through frequency

$$f_{\mathrm{L2}} = \frac{U_1}{D_1/2}. \tag{4.16}$$

Substituting all values, it has $f_{\mathrm{L1}} \approx 0.7747 \mathrm{Hz}$ , $f_{\mathrm{L2}} \approx 11.32 \mathrm{Hz}$.

The frequency of the inertial subrange lies between the energy-containing range and the dissipation range. Therefore, the experimentally measured turbulent fluctuation frequency range of 0~50 Hz aligns with the theoretical estimate of 0.78~55 Hz in magnitude. For the pipes with the diameter $D_2$=1.7cm, one can reasonably expect that the turbulence frequency magnitudes are similar to the measured values at $D_1$=3cm, far below the 1MHz of the acoustic waves generated by the signal generator.

Consequently, turbulent noise cannot resonate with the emitted acoustic waves, failing to satisfy the sound absorption condition described in Howe's theory(vortex sound (2003) and vortex-excited acoustic resonance (Billon, *et al.* 2008; Wang *et al.* 2018).

Additionally, in our experiment, the flow rates under various conditions are all approximately 7.0 L/min. Therefore, with a constant flow rate, the Reynolds number (*Re*) is inversely proportional to the pipe diameter. In Fig.6, *Re* for a pipe diameter of $D_2$=1.7cm is approximately $Re \approx 5094 \times 3/1.7 \approx 8989$, corresponding to the flow conditions at frequencies of 0.9~1.2 MHz in Tables I-III.

### C. Viscous Dissipation

Theoretical studies indicate that the amplitude of acoustic signals inside a pipe decays axially along the pipe due to viscous effects in the form of $\exp(-\alpha z)$, where $z$ is the axial distance and $\alpha$ is the attenuation coefficient of the acoustic wave. Howe (1984) derived the attenuation coefficient for acoustic waves in a pipe with a stationary fluid:

$$\alpha_0 = \frac{l_A}{2AC}\sqrt{\frac{\omega}{2}}\left[\sqrt{\nu} + (\gamma - 1)\sqrt{\chi}\right], \tag{4.17}$$

where $l_A$ is the pipe perimeter, $C$ is the wave speed, $\omega$ is the angular frequency of the signal, $\gamma$ is the ratio of specific heats, and $\chi$ is the thermal diffusivity.

For water at room temperature 18°C, $C \approx 1500\text{m/s}$ , $\gamma \approx 1.000 \sim 1.015$ and $\chi \approx 1.43 \times 10^{-7}\text{m}^2/\text{s}$ . In Fig.6, the pipe diameter $D_2$=1.7cm, $\omega \approx 2\pi \times 10^6\text{Hz}$ . Substituting these values into Equation (4.17) yields:

$$\alpha_0 \approx \frac{1}{D_2 C}\sqrt{2\omega\nu} \approx 0.139/\text{m}. \tag{4.18}$$

The Mach number for the flow is

$$Ma = v_2 / C \approx 3.3 \times 10^{-4}. \tag{4.19}$$

In such condition, it has

$$\alpha_i / \alpha_0 \to 1, (i = 1,2) \tag{4.20}$$

where $\alpha_1, \alpha_2$ are the attenuation coefficients when the acoustic wave propagates in the same and opposite directions as the mean flow, respectively.

In the pipe experiment conducted by Ronneberger and Ahrens (1977), the acoustic wave frequency is 7.35 kHz, and the water flow velocity reaches 50 m/s. The former is far lower than the frequency range (1MHz) in our experiments, while the latter is significantly higher than the flow velocities in our study. According to classical theory, our experiments should exhibit stronger viscous dissipation effects, while the variation in the attenuation coefficient should be very small in turbulent flow.

In Fig.6, the distance between the two transducers is $S$ = 32.0 cm. Under turbulent conditions, more than 40% increase in the acoustic signal compared to the calm state is achieved, so

$$\exp(-\alpha_i \cdot S)/\exp(-\alpha_0 \cdot S) \geq 1.4, \tag{4.21}$$

then, $\alpha_i \leq -1.330/\text{m}$ . Similarly, to achieve a 25% reduction in the acoustic signal under turbulent conditions relative to the static state, the following condition must be satisfied:

$$\exp(-\alpha_i \cdot S)/\exp(-\alpha_0 \cdot S) \leq 0.75. \tag{4.22}$$

So $\alpha_i \geq 1.038/\text{m}$ . These are in serious contradiction with Howe's theory (1984) and the experimental results on attenuation coefficients by Ronneberger and Ahrens (1977).

In addition, viscous dissipation in a pipe primarily concentrates near the solid wall, while in free jet, such viscous dissipation is significantly reduced. Therefore, it cannot explain the variations of acoustic waves by turbulence in jet flow.

**D. Multipath interference**

In our experiments, besides the direct wave, acoustic interference from other propagation paths may also exist. For example, reflected waves that reach the receiver after reflecting off the water surface, pipe wall, pool wall, pool bottom, and refracted/transmitted waves that enter the pipe wall, propagate within it, and then re-radiate back into the pipe or out of it. To examine these multipath acoustic interferences, the following experiments are conducted.

The acoustic characteristic impedances of polyurethane sponge and copper are 0.12 times and 29 times that of water, respectively, so acoustic wave will undergo strong reflection at the surfaces of both materials.

In the pipeline experiments, the characteristic impedance of stainless steel is about 30 times that of water, so very little acoustic energy is transmitted between the water and the pipe wall. Meanwhile, when we wrap the pipe and the transducer with polyurethane sponge, the received signal still fluctuates within its original normal range (the relative change is no greater than 0.3%). In contrast, when the main channel of the pipe is blocked with a copper tube (Fig.11), the amplitude of the received signal drops to below 0.6% of its original value. These indicate that the wave propagating through the main channel is absolutely dominant, while the influence of sound propagating along other paths is extremely small and cannot explain the roughly 30% amplitude variation that occurs under turbulent flow conditions.

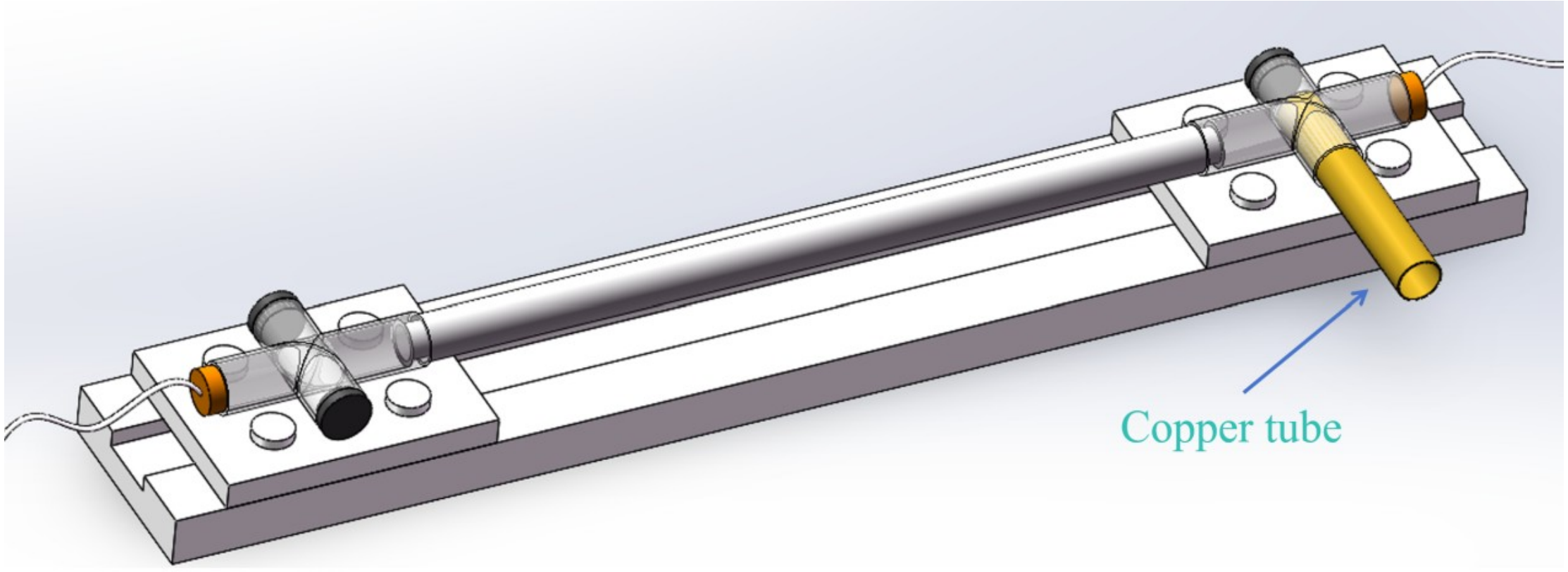

FIG. 11.  Occlusion experiment in a pipe, where the main channel of the pipe is blocked with a copper tube.

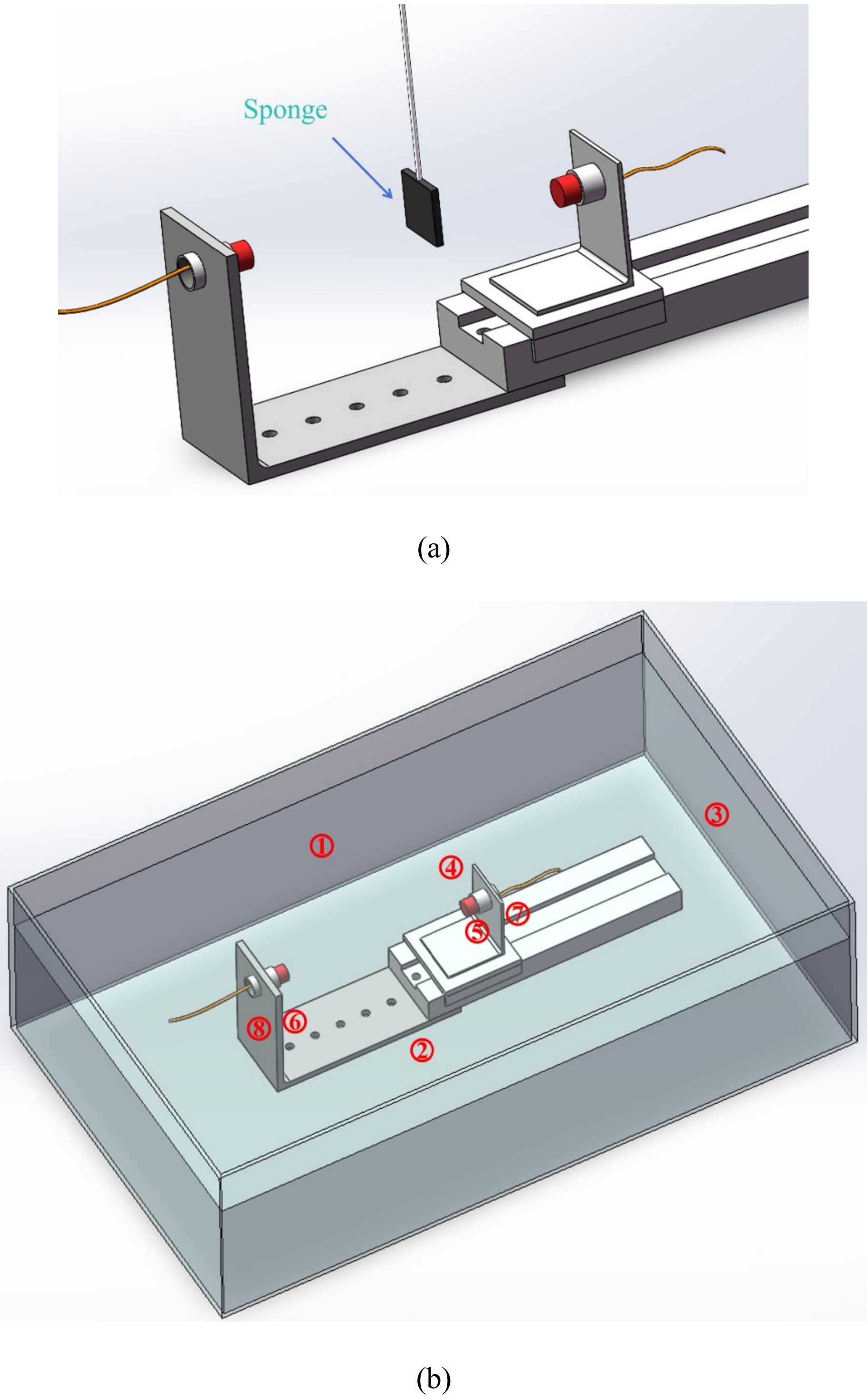

FIG. 12. Occlusion experiment in a free jet: (a) the main channel of the acoustic wave is blocked by sponge; (b) occlusion in other regions, including ① the water surface, ② the pool bottom, ③ the pool wall and ④ the side, ⑤⑥lower parts and ⑦⑧ back of the transducers.

In the free jet experiments, we block the main channel of the acoustic wave between the two facing transducers with polyurethane sponge (Fig.12a), whose size ($5.5\times5.5\times0.9cm^3$) is slightly larger than that of the transducers. As a result, the amplitude of the received signal dropped to 0.8% of its original value. In contrast, when a larger sponge ($15\times10\times0.9cm^3$) is placed at other positions shown in Fig.12b, the amplitude of the received signal changed by less than 0.7% (in a still water pool without the sponge placed, the relative fluctuation of the received signal amplitude is also of this magnitude). This indicates that the wave propagating along the main channel is absolutely dominant, while the interference from waves propagating along other paths is very small. However, the variation in $V_a$ can reach 10%, which also cannot be explained by multipath interference.

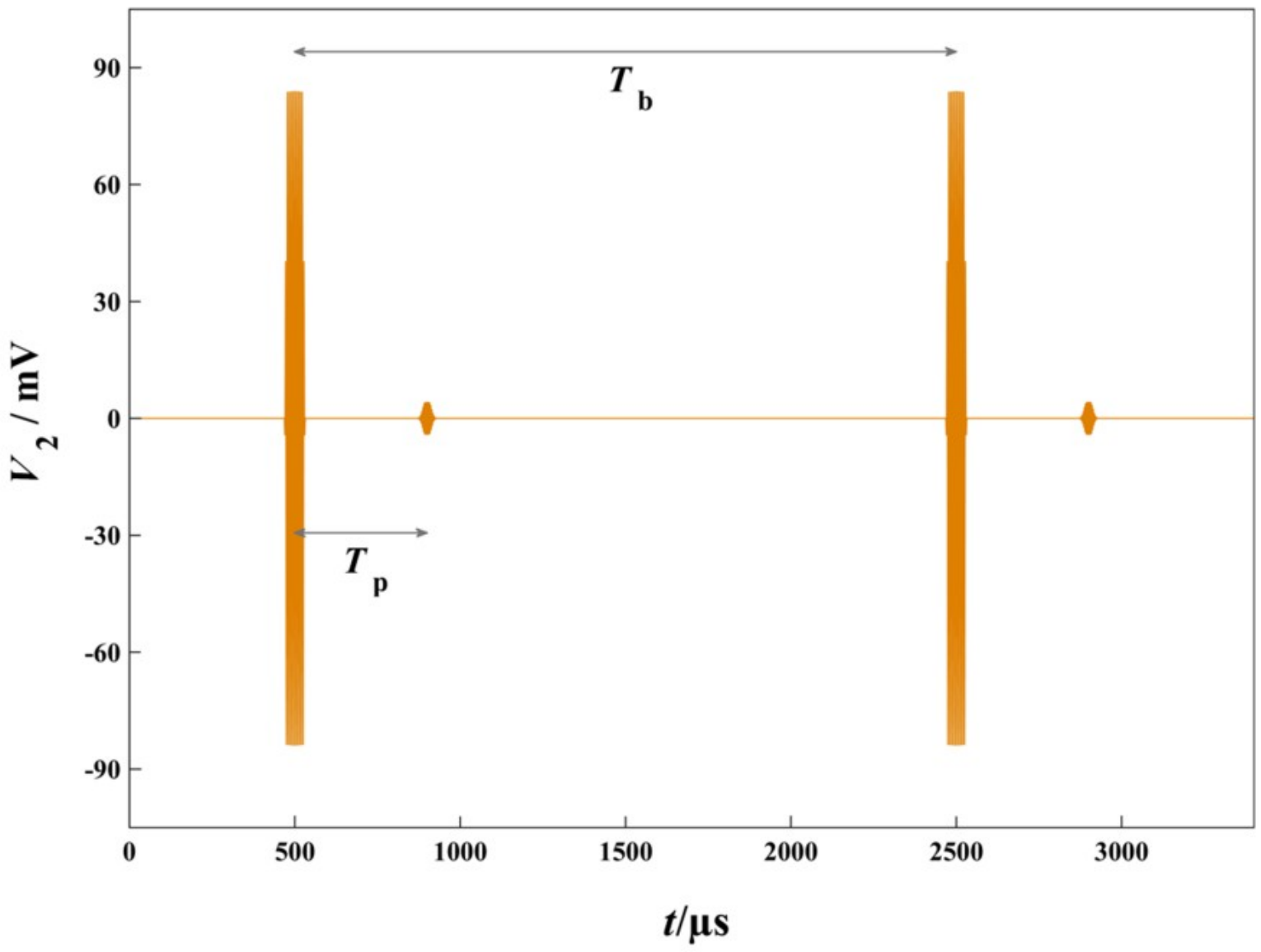

FIG. 13. The received signal in the pipe, including both the direct wave and the reflected wave.

To separate the direct wave from the bypass wave, the signal generator uses a modulation mode, having the transmitter emit $N_p$=40 sine wave pulses at a frequency of around 1 MHz every 2000 microseconds. The received signal in the pipe is shown in figure 13.

It can be seen that the interval between the two pulses with larger amplitudes is $T_b$=2000μs, so they can be identified as direct-wave signals, with an amplitude of $A_d$. Between them, only one type of pulse with a smaller amplitude $A_r$ can be observed; its interval from the preceding direct wave satisfies $T_p \approx 2L_p/C$, where $L_p$ is the distance between the surfaces of the two transducers, and $C$ is the wave speed. So it can be identified as the wave that reaches the receiver again after being reflected twice off the surfaces of the receiver and emitter. As for reflected waves with more reflections, their amplitudes are too small to be distinguished.

The amplitude ratio of the secondary reflected wave to the direct wave ($A_r/A_d$) depends on frequency; for waves around 1 MHz, its value is approximately 1%–10%. Table V shows that turbulence also produces a clear modulation of the direct wave's amplitude.

**Table V**. Modulation of the direct-wave amplitude caused by turbulence.

| No. | Drive Mode | Water Temperature (°C) | Emitter | | Receiver | | | |
|---|---|---|---|---|---|---|---|---|
| | | | Frequency (MHz) | Voltage $V_1$(V) | Static $V_2$(mV) | Turbulence $V_3$(mV) | The relative change $V_a$ | Amplitude ratio $A_r/A_d$ |
| 1 | HHD | 19.2 | 1.05 | 1.5 | 84±1 | 96±1 | ↑,14.3% | 5.0% |
| 2 | HHD | 18.9 | 1.00 | 1.5 | 118±1 | 86±1 | ↓,27.1% | 1.6% |

For the jet flow, the amplitude ratio $A_r/A_d$ is around 1%, and no other obvious traces of bypass waves are observed. This again confirms that the multipath interference is negligible.

**E. Scattering**

Previous literature suggests that when the turbulent flow frequency significantly differs from the acoustic frequency (consistent with our experimental conditions), sound energy is not absorbed by turbulence but rather scattered (1984). Lighthill (1952) has shown that the scattered energy is proportional to the square of the turbulence Mach number. Ford and Meecham (1960) have found that the half-width of the spectrum is proportional to the turbulence Mach number. For turbulent fluctuations in a pipeline without a mean flow in our experiment, the turbulence Mach number $M_t$ is approximately

$$M_t \sim u_* / C \tag{4.23}$$

where $u_*$ is the friction velocity in (4.8). As $O(M_t) \sim 10^{-5}$, the scattering effect caused by turbulence can be neglected.

In addition, a key signature of turbulence-induced acoustic scattering is spectral broadening (1980). However, our experiments never observe this phenomenon (see Fig. 4).

In the experiment by Korman and Beyer (1980), the jet nozzle diameter is 2.54 cm, flow velocity 6.48 m/s, and acoustic frequency 1 MHz. In Fig.2, the corresponding parameters are 0.7 cm nozzle diameter, 3.12 m/s flow velocity, and ~1 MHz acoustic frequency. While the acoustic frequencies are similar in both cases, the flow Reynolds numbers differ by an order of magnitude. In addition, the distance between the turbulent core region and the two transducers is 65 cm in Korman and Beyer (1980), while in our experiment, the distance value is 10~11 cm. Maybe in this condition, the spectral broadening is so small that our oscilloscope is insufficient to detect.

In pipe flow, the confinement by pipe walls prevents acoustic energy from scattering to other spatial locations. In traditional theory, the velocity gradient of a fluid is crucial for scattering (Lighthill, 1952). However, in our experiments, similar variations in acoustic amplitude with no spectral broadening are observed across various flow conditions. This indicates that these phenomena are insensitive to the fluid velocity distribution. The evidence above indicates that the changes in wave amplitude cannot be explained by scattering.

In free jet flow, measurements at multiple azimuth angles reveals no obvious variation in acoustic signals (Table IV). As a comparison, Korman and Beyer (1980) clearly shows that the changes in the spectrum before and after scattering at a scattering angle of 60°. This confirms the absence of directional scattering in the present work, demonstrating that acoustic variation does not originate from scattering mechanisms.

Then we check the Doppler shift effect from a theoretical perspective. Suppose the sound source, receiver, and medium are all moving in rectilinear motion along the same direction, with velocities $U_s$, $U_o$ and $U_m$, respectively. The relationship between the received frequency $\omega$ and the emitted frequency $\omega_0$ satisfies (Pierce, 2019):

$$\omega = \omega_0 \frac{C + U_\mathrm{m} - U_\mathrm{o}}{C + U_\mathrm{m} - U_\mathrm{s}}. \tag{4.23}$$

For the case where the acoustic wave propagates parallel to the mean flow in a pipe, the source and receiver remain stationary $U_\mathrm{s}=U_\mathrm{o}=0$, while the fluid exhibits a mean axial motion . Under these conditions, $\omega = \omega_0$, no Doppler shift occurs.

For the case where the acoustic wave propagates perpendicular to the mean turbulent flow, the fluid velocity component in the transverse direction does not affect the acoustic frequency, and thus no Doppler shift occurs. The above theoretical analysis explains why neither Doppler shift nor spectral broadening is observed in our previous experiments.

**F. Mean flow**

For turbulence under conditions of strong shear, the Rapid-Distortion Theory indicates that the mean flow of turbulence can influence the amplitude and frequency of acoustic waves (Howe, 1995; Bertsch *et al.*, 2012). However, our experiments suggest that the mean flow is not the primary cause of changes in wave amplitude.

From the pump-jet and suction experiments in Fig 6b, it can be seen that: during suction, the receiver’s signal shows no significant difference compared to the static case. A similar result holds when the acoustic waves are perpendicular to the mean flow (Fig. 8).

Additionally, in the experiment of Fig.5, when the valve is closed to halt the mean flow in the pipe, the receiver signal takes a prolonged period to return to its original stable value under static conditions. This demonstrates that turbulent fluctuations (even with zero mean velocity) can induce both amplification and attenuation of acoustic signals. Meanwhile, regardless of whether a mean flow exists, the frequency of the acoustic wave does not change significantly.

These indicate that the mean flow motion is not the cause of acoustic variations.

## G. Temperature

To examine whether the temperature change in water caused by factors such as turbulent viscous dissipation leads to a sufficiently significant alteration in the acoustic wave signal, the following experiment is conducted.

First, we use the water level difference to inject turbulent flow into the pipeline. After a period of time, the valve is closed, the nozzle is removed, leaving only one opening connected to the water in the pool (Fig.14). Then, at intervals of 1 minute, a thermocouple is inserted into the pipeline to measure the water temperature, and the wave amplitude received by the receiver is measured simultaneously. Between two consecutive measurements, the signal generator is turned off to prevent the transmitted acoustic waves from heating the water inside the pipe. Meanwhile, the thermocouple is removed from the pipe.

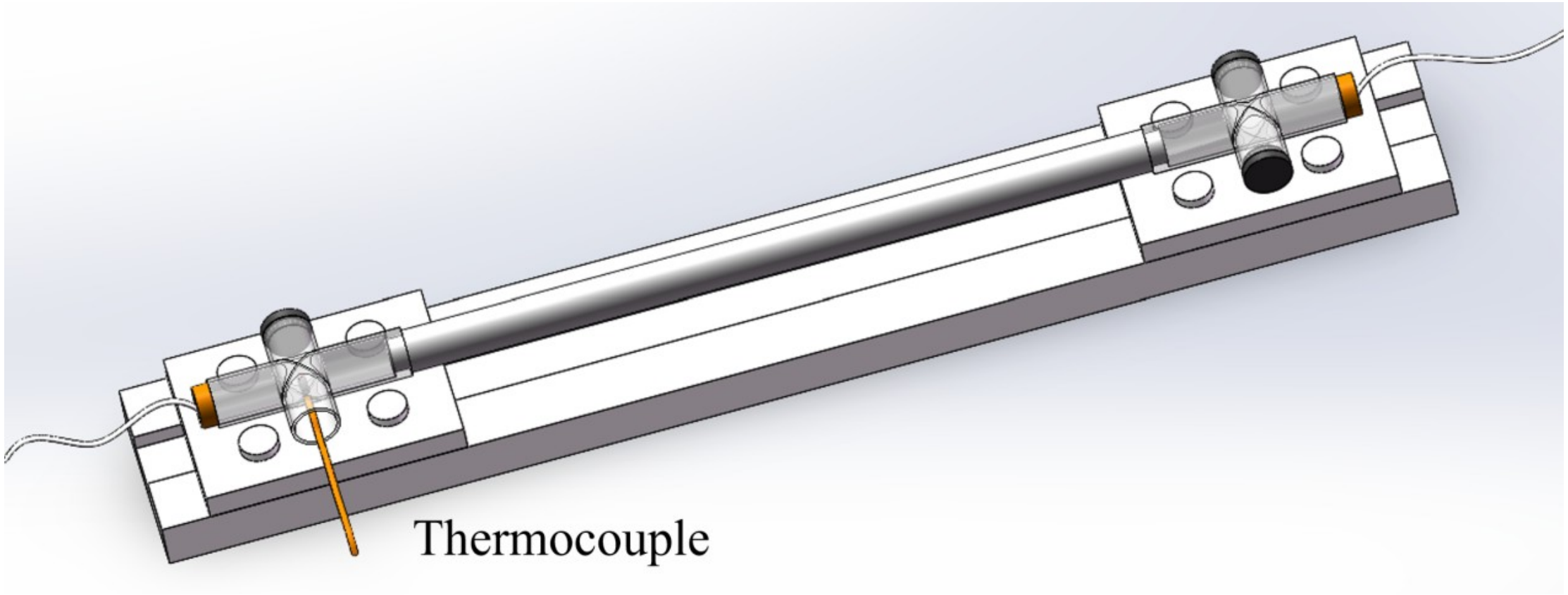

FIG. 14. Schematic diagram of the experiment measuring the wave amplitude and water temperature during the decay of turbulent fluctuations inside the pipeline.

**Table VI** The values of wave amplitude and water temperature during the decay of turbulent fluctuations. The transmitted signal is set to 1.112MHz at an excitation voltage of $V_1$ = 0.5 V while the flow driven by a water level difference.

| Time (Min) | Water temperature (°C) | The received signal $V_3$ (mV) |
|---|---|---|
| 0 | 24.3±0.05 | 30.8±0.1 |
| 1 | 23.4±0.05 | 27.8±0.1 |
| 2 | 23.4±0.05 | 26.8±0.1 |
| 3 | 23.4±0.05 | 26.2±0.1 |
| 4 | 23.4±0.05 | 25.6±0.1 |
| 5 | 23.4±0.05 | 25.3±0.1 |
| 6 | 24.3±0.05 | 25.0±0.1 |

In Table VI, the wave amplitude decreases from an initial value of 30.8 mV to 25.0 mV after 6 minutes, while the temperature remained the same before and after. During this period, the water temperature fluctuates by approximately 0.1°C, while the wave amplitude steadily declines. Similar results are observed for flow driven by a pump.

In pipe flows, a temperature boundary layer may exist between the fluid and the wall. However, in the free jets, no temperature boundary layer is present in turbulent region, yet turbulence can still significantly affect the amplitude of acoustic waves.

These observations strongly indicate that temperature change is not the primary factor influencing the amplitude of acoustic waves.

## H. Mechanical vibration

To what extent do turbulent fluctuation and mechanical vibrations from the acoustic emitter affect the signals receiver? In the pipeline experiments, the pipes, guide rails, and sliders are all made of stainless steel, while the connectors are made of PVC. The transducers are securely wrapped with tape and fixed inside short PVC pipes. In the free jet experiments, the PVC pipes are also connected to stainless steel supports. Since the characteristic impedance of stainless steel is approximately 14 times that of PVC, the connection between the two can effectively isolate vibrations.

Additionally, when we seal both the inlet and outlet of the water flow in the pipeline with plugs, then use a nozzle to direct a jet impact on various parts of the pipeline, varying the angles and even making the nozzle contact the pipeline (Fig.15). The received signal values under these conditions show no difference compared to those in stationary water. For the free jet, similar experiments are conducted and the results are the same: as long as the turbulent flow does not pass through the acoustic propagation region, the jet impact and contact on the transducer support structure alone do not cause any fundamental changes in the received signal. This also rules out the influence of structural vibration.

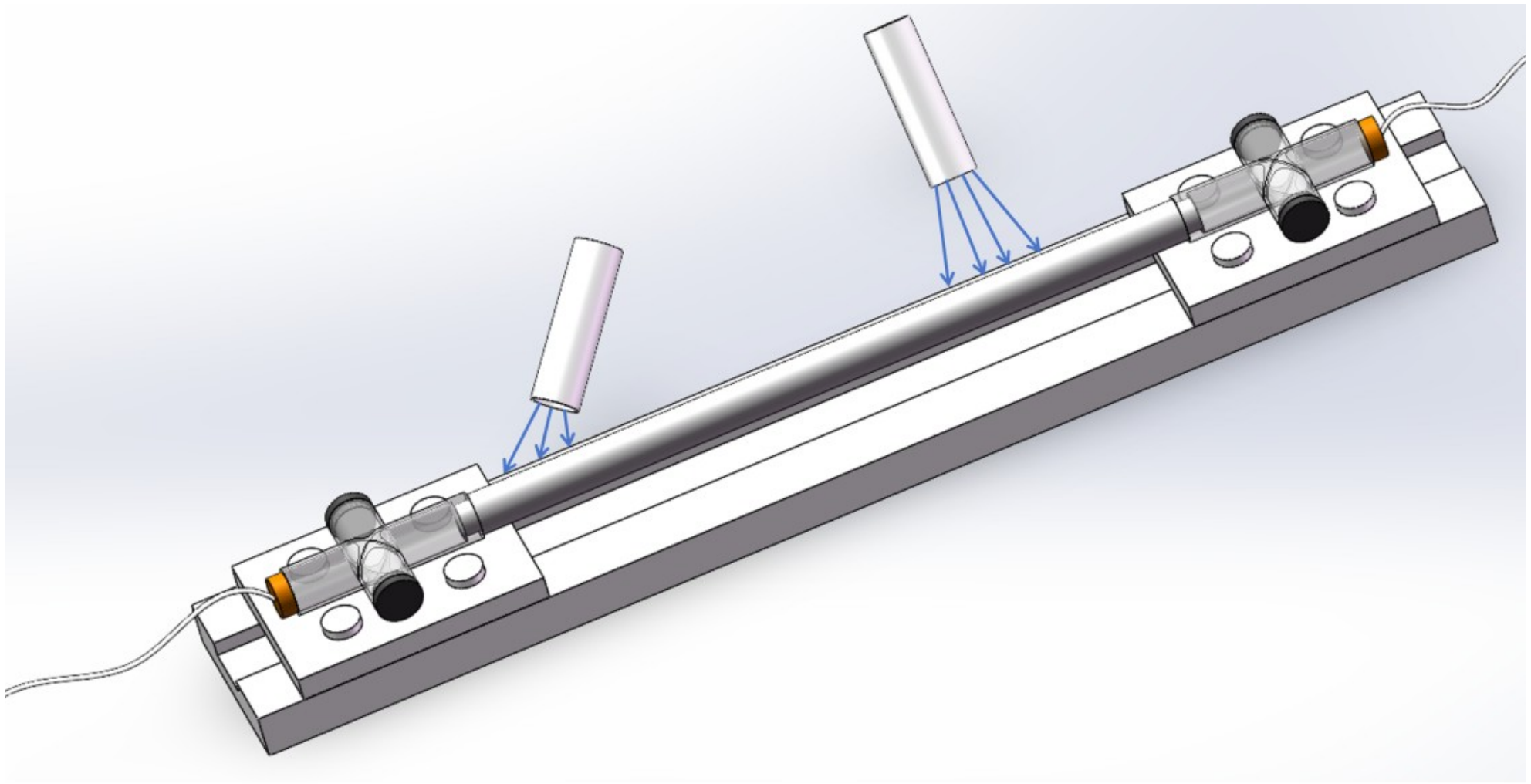

FIG. 15. Schematic diagram of the experiment on jet impact and contact with the transducer support structure.

When the emission signal is turned off, the received signal immediately drops to the noise level. This indicates that turbulence itself cannot generate acoustic waves of this frequency through structural vibrations to affect the receiver.

In summary, one can infer that the influence of mechanical vibration is not crucial.

## I. Interactions between different modes

In our experiments, does energy transfer exist between different modes? First, we consider the modes with different frequencies. When the signal generator emits a square wave with a fundamental frequency of 0.25 MHz, several distinct peaks could be observed in the spectrum of the receiver within the range of (0.25~16) MHz. So the transducer can effectively detect sinusoidal waves within this range.

In our experiments, the signal generator emits a single-frequency sinusoidal wave. In the received signal, only the wave at the same frequency has its amplitude modulated by turbulence, while waves at other frequencies show no significant change. These indicate that energy is not transferred between the modes with different frequencies!

Next, the modes with the same frequency are examined. According to acoustic theory, when sound waves propagate in a steel pipe, if the frequency exceeds the cutoff frequency $f_c = 1.841\frac{c}{\pi D}$, higher-order modes may exist in addition to the fundamental plane wave mode. They are generated due to non-ideal factors such as turbulent fluctuations, velocity boundary layers caused by viscosity, non-uniformity of vibration, and finite duct length. The boundary layer thickness is $\beta = \sqrt{\nu/2\pi f}$, where $\nu$ is the kinematic viscosity of the fluid.

In our experiment, when $D$=1.7cm, $f_c$=51.7kHz $<< f = 1\text{MHz}$, so high-order modes can exist. However, the fluctuating velocity of turbulence is much smaller than the speed of sound. The boundary layer thickness $\beta \approx 0.4\times10^{-6}\,\text{m}$ at $f = 1\text{MHz}$, which is far less than the wavelength $\lambda \approx 1.5\text{mm}$. The transducers emit acoustic waves in the form of a circular piston array. Meanwhile, the pipe length is also much greater than the diameter. Therefore, the influence of non-ideal factors is very small, and the amplitude of the higher-order modes is also far less than that of the plane wave. The receiver reflects the average sound pressure on the interface, thus primarily reflecting the amplitude of the plane wave. The energy interaction between higher-order modes and the fundamental mode cannot cause such a significant change in the received signal.

Most importantly, for the free jet flow in Sec. III B, acoustic waves do not have discrete multiple modes. However, turbulence still alters the wave amplitude, which cannot be explained by the energy redistribution among modes.

## V. CONCLUSION

This work experimentally examines the effects of turbulence on hydroacoustic wave, focusing on two distinct turbulent flow configurations: pipe flow and free jet. The flows are driven by either a high-pressure pump or hydraulic head difference, with hydroacoustic transducers operating in the 0.9~1.2 MHz frequency range. The results demonstrate that acoustic waves can undergo significant attenuation or amplification when propagating through turbulence, at frequencies far exceeding the characteristic turbulent frequencies. The amplification factor depends on the wave frequency rather than its amplitude. Notably, no spectral broadening is observed in any of the tested conditions.

In pipe flow experiments, both parallel and perpendicular orientations of acoustic waves relative to the mean flow are investigated. Great signal variation (exceeding 40% in magnitude) occurred, suggesting strong turbulence-wave interaction.The total amplification factor of acoustic waves in the entire pipeline is equal to the product of the amplification factors in each section of the pipeline.

After closing the flow control valve, the mean flow motion ceases, the receiver signal exhibits a finite relaxation period before recovering to its stationary-state value, while both amplification and attenuation occur during the decay of turbulence. Laminar flow produces no measurable alteration of acoustic waves. The primary mechanism for acoustic amplitude modulation stems from turbulent fluctuations rather than mean flow.

For jet-induced turbulence, the acoustic wave propagates perpendicular to the flow direction, and the received signal also exhibit amplification or attenuation. When the receiver is placed at different azimuth angles, the occurrence of turbulence does not cause significant changes in the received signal.

For each scenario in this experiment, the amplitude of acoustic waves across all frequency bands either uniformly decreases or increases under the influence of turbulence. There are no instances where some frequency bands decreases while others increases simultaneously, and no new spectral component appears. When the wave propagation direction is parallel or perpendicular to the mean flow, no spectral broadening or Doppler shift are observed. When the transmitter is turned off, the received signal drops to the noise level—regardless of whether the water is stationary, in laminar flow, or turbulent—except for the turbulence-induced frequency. These findings indicate that turbulence only attenuates or amplifies the incident acoustic wave, but does not generate new frequency components.

## VI. DISCUSSION

The observed experimental phenomena cannot be explained by conventional mechanisms such as bubbles, resonance, scattering, multipath interference, or viscous dissipation. These hints suggest that there may be a new mechanism for the interaction between turbulence and acoustic waves.

In addition, this paper does not provide measurements of various details within the turbulence, such as the spatial distribution of the mean flow and turbulent kinetic energy. The reason can be explained as follows. Our experiments include pipe flow, free jet flow, and turbulent fluctuation decay without a mean flow. The flow driving methods are divided into two types, pipes come in various diameters and lengths, and the mean flow is parallel or perpendicular to the direction of wave propagation. The details of these flows differ, yet all can alter the wave amplitude.

As preliminary work, this study aims to reveal that the fundamental cause of the change in wave amplitude is turbulent fluctuation, which is independent of factors such as the type of mean flow and the direction of its velocity. The investigation of the impact of turbulent kinetic energy and its distribution on the amplification factor will be addressed in subsequent work.

**APPENDIX**

Here, the key materials and instruments used in the experiments are listed.

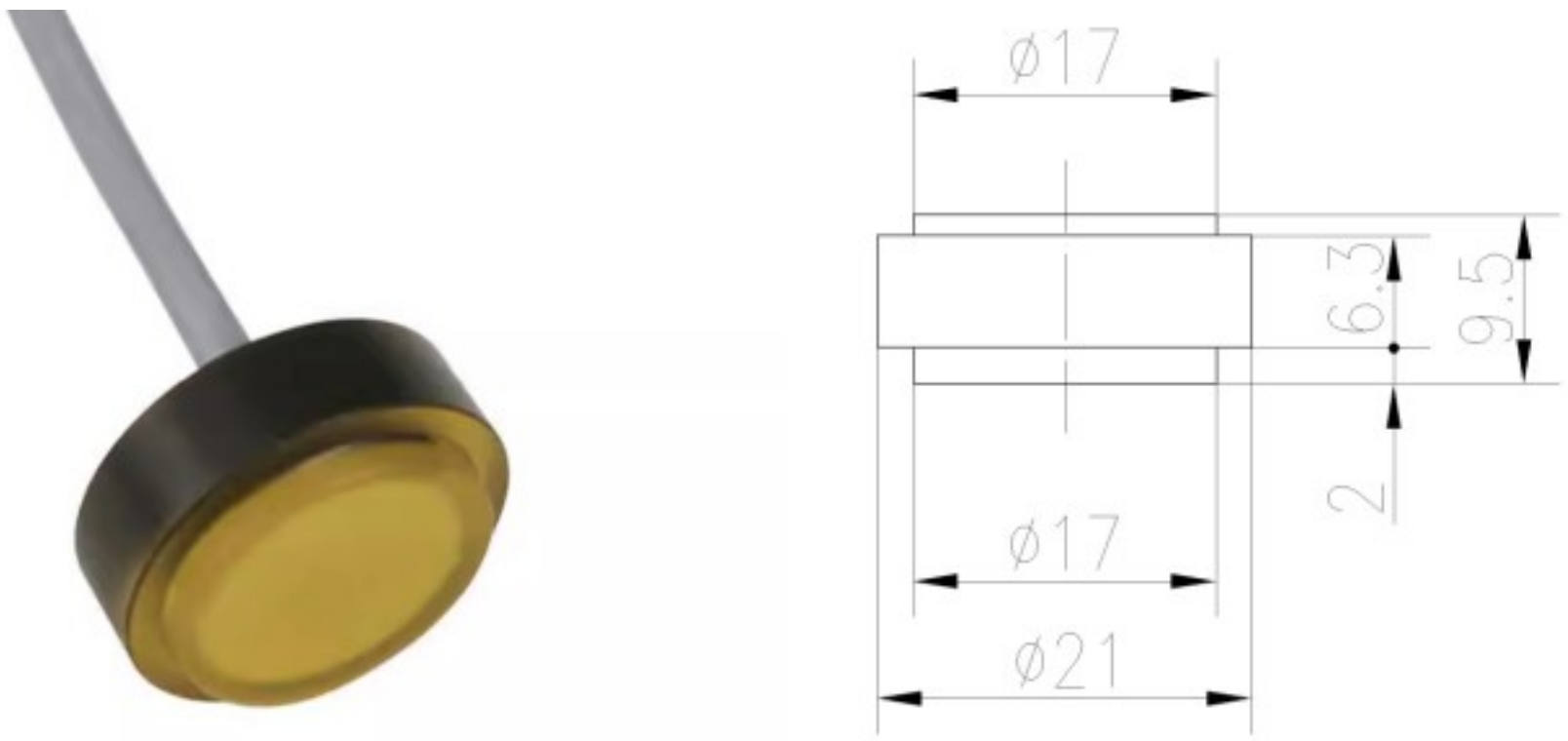

(a) (b)

FIG. 16. Hydroacoustic transducer with the fundamental frequency of 1MHz: (a)Physical diagram; (b)dimension diagram, in which the unit is mm.

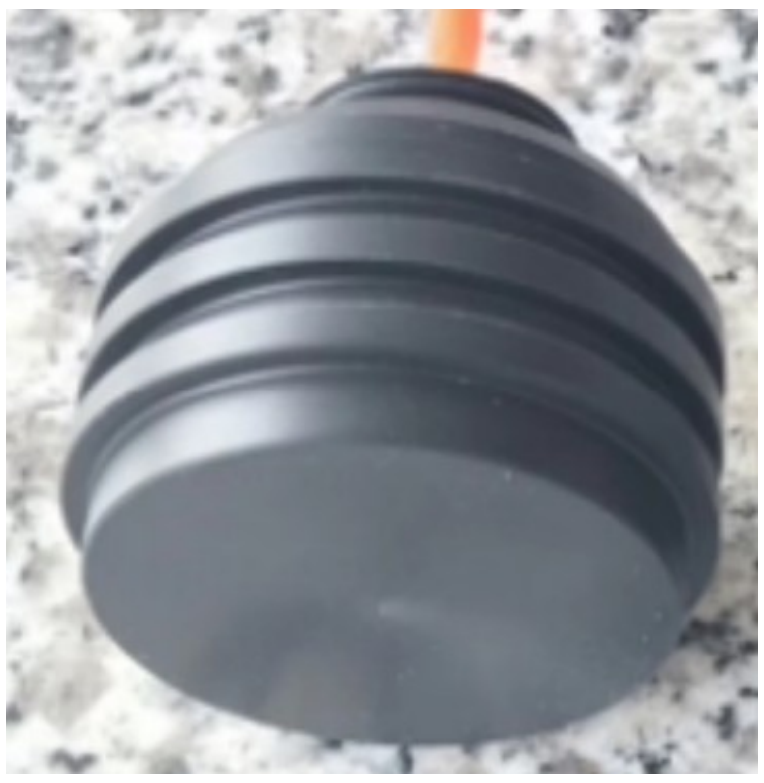

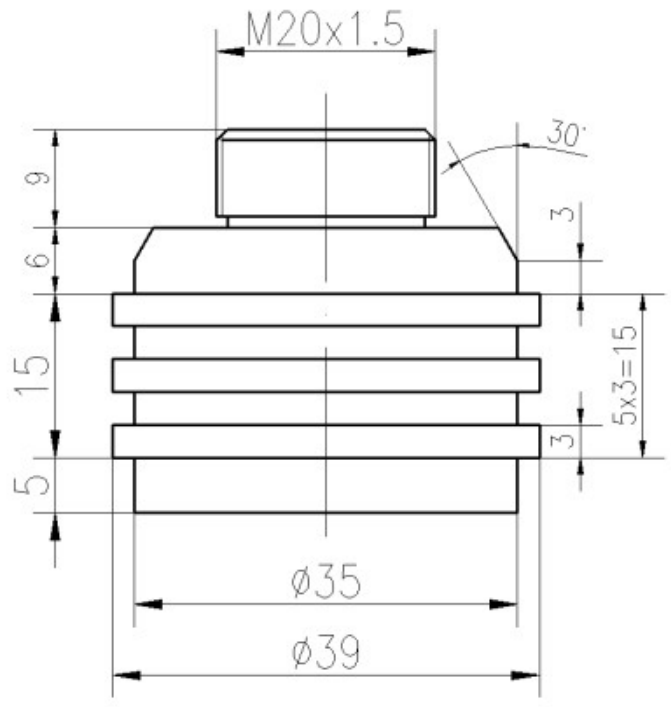

(a) (b)

FIG. 17. Hydroacoustic transducer with the fundamental frequency of 7KHz: (a)Physical diagram; (b)dimension diagram.

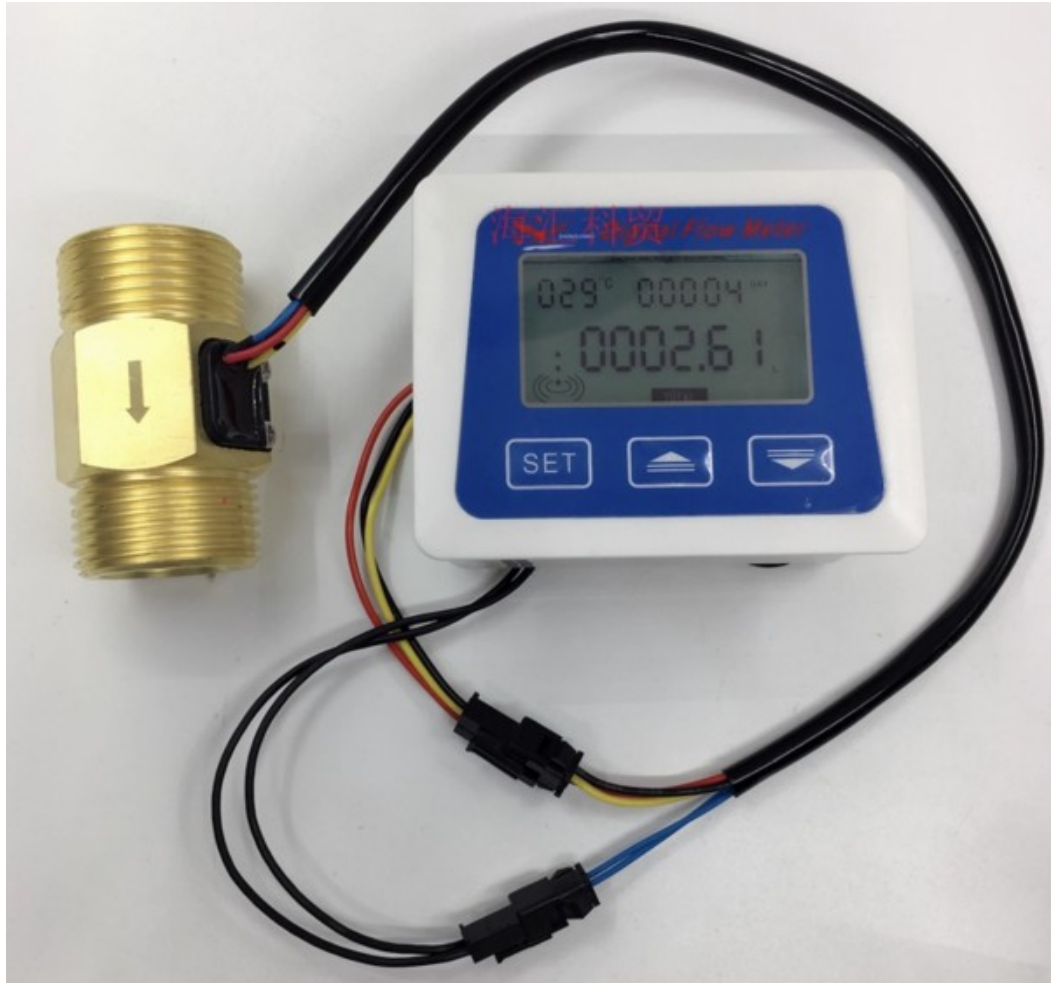

FIG. 17. Turbine flowmeter.

Fig.18 shows the turbine flowmeter, Table VII summarizes other major instruments, and Table VIII provides the parameters annotated in the Figs.1, 2.

**Table VII** Instruments

| Instrument name | Model | Manufacturer |
|---|---|---|
| Signal Generator | 2040H | Victor |
| Oscilloscope | MDO-2202ES | GW Instek |
| High-pressure pump | — | Zengyuan Electromechanical |
| Hydroacoustic transducer | DYW-1M-01I<br>DYW-7-ND | An Buleila Ultrasound |
| Manual Linear Stage | 40A-35 | Hengyu Laser Equipment |
| Turbine flowmeter | — | Haihui Trading |
| PVC Connectors | — | Baisheng Pipeline |
| Stainless steel pipe | 304 | Qing-zhi |
| Linear Guide Rails | SGR10 | Xinrui Bearing |

**Table VIII** The parameters in Figs.1, 2.

| No. | Flow | Transducer's fundamental frequency/MHz | Parameters |
|---|---|---|---|
| 1 | Pipe flow | 1 | $S$=32cm, $D$=1.7cm, |
| 2 | Pipe flow | 0.007 | $S$=23cm, $D$=3.0cm, |
| 3 | Free jet | 1 | $h_1$=14.5cm, $h_2$=16cm, $h_3$=10cm,<br>$h_4$=4.5cm, $h_5$=5cm. |

## DATA AVAILABILITY

The authors declare that the data supporting the findings of this study are available within the paper and its supplementary information files. Source data are available upon reasonable request.

## DECLARATION OF INTERESTS

The authors declare no conflicts to disclose.

## AUTHOR CONTRIBUTIONS

Kai-Xin Hu made substantial contributions to the conception of the work, wrote the paper for important intellectual content and approved the final version to be published. He is accountable for all aspects of the work in ensuring that questions related to the accuracy or integrity of any part of the work are appropriately investigated and resolved. Yue-Jin Hu designed and fabricated the experimental apparatus, and conducted experimental measurements jointly with Kai-Xin Hu.

## ACKNOWLEDGMENTS

This work has been supported by the National Natural Science Foundation of China (No.12372247), Zhejiang Provincial Natural Science Foundation (No.LZ25A020009) and the China Manned Space Engineering Application Program—China Space Station Experiment Project (No. TGMTYY14025).